\documentclass{aa}
\usepackage{txfonts}
\usepackage{natbib}
\usepackage{graphicx}
\bibpunct{(}{)}{;}{a}{}{,}

\begin{document}

\title{Effective area calibration of the Reflection
Grating Spectrometers of XMM-Newton. I. X-ray spectroscopy of the Crab nebula}

\author{J.S. Kaastra\inst{1,2}
 \and C.P. de Vries\inst{1}
 \and E. Costantini\inst{1,2}
 \and J.W.A. den Herder\inst{1}
}

\offprints{J.S. Kaastra}

\institute{SRON Netherlands Institute for Space Research, Sorbonnelaan 2,
           3584 CA Utrecht, the Netherlands 
	   \and
	   Sterrenkundig Instituut, Universiteit Utrecht, 
P.O. Box 80000, 3508 TA Utrecht, The Netherlands
	 }

\date{\today}

\abstract
{ The Crab nebula and pulsar have been widely used as a calibration source 
for X-ray
instruments. The in-flight effective area calibration of the Reflection
Grating Spectrometers (RGS) of XMM-Newton depend upon the availability
of reliable calibration sources. }
{ We investigate how the absolute effective area calibration of RGS can
be obtained using Crab as a standard candle.} 
{ We have analysed RGS observations of the Crab using different instrument
configurations and spatial offsets, and made use of previous determinations
of the continuum spectrum of the nebula plus pulsar. Due to the high spectral
resolution of the RGS, we resolve the main absorption edges and detect the strong
1s--2p absorption lines of neutral oxygen. }
{ We get an excellent fit to the Crab spectrum using this fixed continuum
and the absorption spectrum determined by RGS. We get accurate column densities
for the neutral atoms of H, N, O, Ne, Mg, and Fe, as well as a clear detection
of \ion{Fe}{ii} and firm upper limits for other ions. Our data are in good
agreement with earlier optical and UV spectroscopic measurements of
some of these ions. We find solar abundances for N and O, while
Ne is overabundant by a factor of 1.7 and Fe is underabundant by a factor of
0.8. We confirm that there is less dust in the line of sight compared to
the prediction based on the absorption column. 
Our spectra suggest a more prominent role of
ferric iron in the dust compared to ferrous iron. } 
{ Our high-resolution observations confirm that Crab can be used as
an X-ray calibration source. RGS spectra have determined the absorption
spectrum towards Crab with unprecedented detail.}


\titlerunning{X-ray spectroscopy of the Crab nebula}
\authorrunning{J.S. Kaastra et al.}

\maketitle

\section{Introduction}

The \object{Crab} nebula and pulsar have been widely used as a calibration
source for X-ray instruments, starting already in the early days of X-ray
astronomy. The main advantage of Crab is that it is one of the brightest steady
X-ray sources on the sky and that it has a simple power-law spectrum.

Later, more complexities of the spectrum have become clear. For instance, recent
imaging data (XMM-Newton: \citealt{willingale2001}; Chandra:
\citealt{weisskopf2000,mori2004}) show a complex spectral and spatial structure
with the components such as pulsar, torus, jet, dust-scattering halo, etc. Also,
the spectral shape of the pulsar and nebula are different, and both show
spectral breaks at higher energies (e.g., \citealt{kuiper2001}).

A complicating factor is that the interstellar absorption becomes very high at
low energies. For many instruments, the uncertainties in the low-energy
calibration of the instruments is formally compensated for by adjusting the
interstellar absorption column density to get a best fit. This procedure,
however, introduces high systematic errors both in the derived effective area
at low energies, as well as in the derived interstellar absorption. This issue
is illustrated by \citet{kirsch2005}, who compare a large number of current
and past X-ray instruments that have observed Crab.

There is still an urgent need for accurate calibration of X-ray instruments in
space. First, an accurate knowledge of the relative effective area is needed for
sources with complex intrinsic continua, for example active galactic nuclei with
warm absorbers, relativistically broadened iron or oxygen lines, soft spectral
components and hard reflection components. Furthermore, an accurate knowledge of
the absolute effective area is important when X-ray data are to be matched to
optical/UV data (for example for isolated neutron stars), or where the absolute
flux has cosmological implications. A good example of the last issue are
clusters of galaxies; their absolute flux determines their absolute hot gas mass
as well as the inferred baryonic fraction.

The problem with most previous and current X-ray instruments that have observed
Crab is their relatively poor spectral resolution at the low energies where
interstellar absorption is important. The situation has improved considerably by
the advent of high-resolution grating spectrometers. With high spectral
resolution, it is possible to resolve the absorption edges and absorption lines,
thereby constraining the interstellar absorption. With the Chandra Low Energy
Transmission Grating Spectrometer (LETGS) it has been possible to obtain a high
resolution spectrum of the Crab pulsar \citep{tennant2001}, but the nebula
containing most of the X-ray flux is too extended for the LETGS to obtain a high
resolution spectrum. On the other hand, the Reflection Grating Spectrometers
(RGS) of XMM-Newton \citep{denherder2001} are well-suited to observe the full
Crab nebula. 

From an astrophysical point of view, these Crab spectra offer us a unique
opportunity to study the interstellar medium. For instance, the determination of
the interstellar abundances from optical and UV spectroscopy is not
straightforward, as only the gaseous components yield absorption lines.
Moreover, these absorption lines are often heavily saturated, thereby making an
accurate determination of ionic column densities difficult. Furthermore,
depletion of free atoms onto dust complicates the situation considerably. In
X-ray spectra essentially all atoms, whether in gaseous or molecular form,
contribute to the X-ray opacity, and the correction factors for differences in
dust and gaseous opacity are not very high. Moreover, high-resolution spectra
near the dominant absorption edges can help to disentangle the gaseous from the
dusty or molecular phases, and to give clues on the chemical composition of the
dust.

In this paper we analyse a set of observations of the Crab nebula, obtained with
the RGS of XMM-Newton. These data were taken for the purpose of effective area
calibration. The paper is laid out as follows. In Sect.~\ref{sect:intrinsic} we
discuss the intrinsic (unabsorbed) continuum spectrum of Crab as derived from
other instruments. In Sect.~\ref{sect:data} we describe the basic data
extraction and non-standard instrumental corrections that we need to apply.
Sect.~\ref{sect:analysis} then presents the results of our spectral analysis. In
Sect.~\ref{sect:discussion} we discuss our results on the interstellar
absorption and the composition of the interstellar medium towards Crab. In 
Sect.~\ref{sect:rgs} we discuss the implications for the calibration of the RGS
and give a breakdown of the systematic uncertainties involved in this
calibration. Our conclusions are given in Sect.~\ref{sect:conclusions}.

This paper is the first of a series of three papers. In paper II \citep{paper2}
we present a new analysis of the X-ray model spectra of the white dwarfs
Sirius~B and HZ~43. We show in that paper how these models can be used to
calibrate the long wavelength absolute effective area of the LETGS of Chandra
with an accuracy of a few percent. In paper III \citep{paper3} we combine these
results based on white dwarfs with our present results on Crab using blazar
spectra taken simultaneously with the RGS and LETGS. This leads to a refinement
of the calibration presented in the present paper, and allows to determine the
absolute flux of Crab within a few percent uncertainty. 

\section{The intrinsic spectrum of the Crab nebula and
pulsar\label{sect:intrinsic}}

\subsection{The Crab nebula as a calibration source}

have studied the Crab nebula as a potential calibration source
for X-ray astronomy. They measured Crab in 1970 using a rocket experiment with
proportional counters between 1.5--10~keV and scintillation counters between
5--70~keV. The measured spectrum (nebula $+$ pulsar) has a power-law with photon
index $\Gamma = 2.10\pm 0.03$ and normalisation $N$ at 1~keV of $9.7\pm 1.0$ (in
units of $10^4$~photons\,m$^{-2}$\,s$^{-1}$\,keV$^{-1}$, used throughout this
paper for the Crab normalisation). Apparently no correction for Galactic
absorption at the lowest energies was made.

If we fit the data as presented by Toor \& Seward, we find instead
$\Gamma=2.13\pm0.05$ and $N=10.3\pm0.9$, i.e. a slightly steeper power-law.
Correcting their data for Galactic absorption using simply \citet{morrison1983}
absorption cross sections and $N_{\mathrm H}=3.45\times 10^{25}$~m$^{-2}$
\citep{schattenburg1986}, we obtain a 20~\% higher flux at 1.7~keV, the lowest
energy measured by Toor \& Seward. This results in $\Gamma=2.18\pm0.05$ and
$N=11.5\pm1.0$, i.e. an increase in $\Gamma$ of 0.05.

\begin{figure}
\resizebox{\hsize}{!}{\includegraphics[angle=-90]{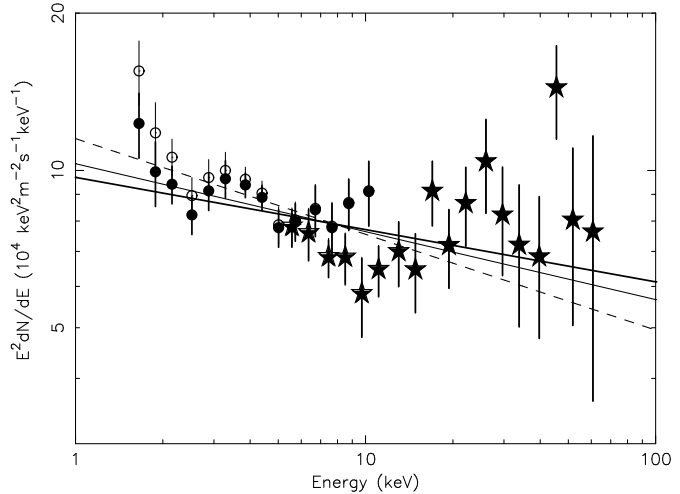}}
\caption{Measurements of the Crab nebula by \citet{toor1974}.
Circles: proportional counter; stars: scintillation counter; filled symbols:
data from Fig.~3 of Toor \& Seward; open symbols: same data but corrected
for Galactic absorption. Thick solid line: power-law model of Toor \& Seward;
thin solid line: our power-law fit to the same data; dashed line: power-law
fit to absorption corrected data.}
\label{fig:toor}
\end{figure}

Fig.~\ref{fig:toor} shows a comparison between these fits and the raw and
absorption-corrected data of Toor \& Seward. It is evident that at any energy
the uncertainty in the spectrum is at least 10~\%. Also note the discrepancy
between both instruments near 10~keV: the scintillation counter measures about
13~\% less flux in the common 5--11~keV band.

From the above discussion it is clear that we need to pay more attention to the
way the spectral parameters of Crab are derived and to their statistical and
systematic uncertainties. As it is known that the pulsar and the nebula have a
different spectrum, we also have to derive both model spectra separately, even
though our RGS observations have insufficient spatial and time resolution to
distinguish both.

Since the early work of \citet{toor1974}, many measurements of the Crab
spectrum using different instruments have been made, see for instance
\citet{kirsch2005} for an overview. Not all of these measurements are suitable
for our purpose. Most instruments operating only in the energy band below 10~keV
have hard times in determining reliable spectral indices. This is because at low
energies the effects of 1) low spectral resolution, 2) complicated instrumental
effective area structures like filter edges etc., 3) a complex redistribution
function and 4) the strong impact of interstellar absorption all conspire to
make an accurate flux measurement at low energies and thereby a reliable photon
index in the 1--10~keV band a hard to determine quantity. At higher energies all
these effects are less important. For this reason, the data set that we use for
the spectral parameters of Crab is taken from the analysis of BeppoSAX and other
data as presented by \citet{kuiper2001}. These are based on a careful analysis
of the spectrum in a very broad band, separating the pulsar and nebula
component, and although there still is a 10~\% uncertainty in the absolute flux
scale of those data, the spectral shape is very well determined, as elaborated
below.

\subsection{The X-ray spectrum of the Crab pulsar\label{sect:pulsarmodel}}

\citet[K01 hereafter]{kuiper2001} have studied Crab extensively using data from
BeppoSAX, Comptel and EGRET. From their paper, we find the best spectrum of the
pulsar to be given by
\begin{equation}
F_p = 726E^{-1.276}e^{-0.074x^2}+1464E^{-1.165}e^{-0.159x^2}+2021E^{-2.022},
\label{eqn:pulsar}
\end{equation}
where $E$ is the photon energy in keV, $x\equiv \ln E$ and $F_p(E)$ the spectrum
in photons\,m$^{-2}$\,s$^{-1}$\,keV$^{-1}$. This spectrum needs to be corrected
for absorption by the intervening ISM, for which K01 used \citet{morrison1983}
cross sections with $N_{\mathrm H}=3.61\times 10^{25}$~m$^{-2}$.

\subsection{The neutron star (unpulsed) emission}

The spectrum of the pulsar as described by K01 only contains the pulsed
emission. There is also steady emission. \citet{tennant2001} estimate the zeroth
order count rate of the steady neutron star emission as measured by the Chandra
LETGS as 0.19~counts\,s$^{-1}$. For comparison, the total (dead-time corrected)
zeroth order count rate is 8.16~counts\,s$^{-1}$, hence the steady emission is
only 2.3~\% of the total emission. \citet{weisskopf2004} estimated the steady
neutron star emission from the LETGS spectrum. For a distance of 2~kpc and a
radius of 15.6~km, they find a 2$\sigma$ upper limit to the temperature of
0.159~keV. Such a spectrum produces only 0.3--1~\% of the total point source
emission over the full LETGS range, 1~\% being reached around 0.8~keV. Hence we
conclude that the contribution of the unpulsed emission can be neglected for the
present purpose.

\subsection{The X-ray spectrum of the Crab nebula\label{sect:nebulamodel}}

A simple power-law approximation for the nebula or the pulsar over the full
X-ray and $\gamma$-ray energy range is inadequate. At higher energies, the
spectrum of Crab softens. This was shown by K01 using Comptel data and by
\citet{ling2003} using BATSE data. The break of about 0.3 in photon index occurs
around 100~keV. Ling \& Wheaton give an average photon index of $2.1\pm 0.1$
below the break at $120\pm$40~keV and $\Gamma = 2.4\pm 0.2$ above the break.
Therefore, we will limit the range of validity of any power-law approximation to
$E<100$~keV.

\begin{figure}
\resizebox{\hsize}{!}{\includegraphics[angle=-90]{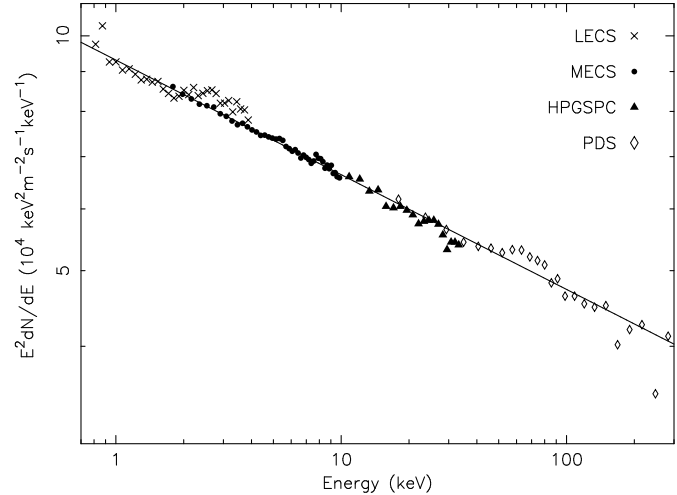}}
\caption{BeppoSAX measurements of the Crab nebula from \citet{kuiper2001}. The
pulsar has been subtracted and the fluxes have been corrected for Galactic
absorption as described in the text. LECS, HPGSPC and PDS data have been divided
by 0.93, 1.01 and 0.87, respectively. For clarity, error bars (all lower than
the plot symbols) have been omitted. The solid line indicates the best fit
power-law excluding LECS) with $\Gamma=2.147\pm 0.002$ and $N=9.31\pm 0.05$. }
\label{fig:crab_k_nebula}
\end{figure}

K01 also studied the spectrum of the nebula only (pulsar subtracted). These data
are available at http://www.sron.nl/divisions/hea/kuiper/data.html. From their
data, we find the best fit power-law based upon BeppoSAX MECS data only to be
$\Gamma=2.146\pm 0.003$ and $N=9.31\pm 0.05$ (no systematic uncertainty
included).  We have fitted all BeppoSAX data as given by K01 (MECS, HPGSPC and
PDS, excluding LECS) in the 1--100~keV band using a simple power-law, and
scaling the model for HPGSPC by a factor of 1.01 and PDS by a factor of 0.87
with respect to MECS, following K01. We excluded LECS because it is more
sensitive to Galactic absorption and also because its spatial and spectral
resolution decreases rapidly for lower energies, making it more prone to
possible systematic uncertainties. The best fit power-law has $\Gamma=2.147\pm
0.002$ and $N=9.31\pm 0.05$, in excellent agreement with the fit to MECS only.
We show the fit in Fig.~\ref{fig:crab_k_nebula}. Because the residuals are
dominated by small remaining systematic variations of less than 1, 2 and 4~\%
for MECS, HPGSPC and PDS, respectively, we have used an unweighted fit through
these data points. Hence our adopted model for the nebula is given by 
\begin{equation}  
F_n = 9.31\times 10^4 E^{-2.147} 
\label{eqn:nebula}
\end{equation}  
where again $E$ is the photon energy in keV and $F_n(E)$ the nebular spectrum in
photons\,m$^{-2}$\,s$^{-1}$\,keV$^{-1}$.

\begin{figure}
\resizebox{\hsize}{!}{\includegraphics[angle=-90]{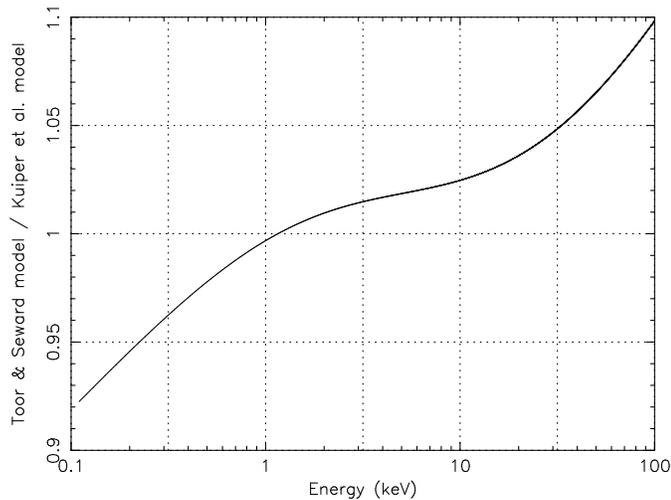}}
\caption{Comparison between the model spectra for the full Crab nebula and
pulsar. Shown is the ratio between the power-law model of
\citet{toor1974} and the more complicated model of \citet{kuiper2001}.}
\label{fig:ts_k}
\end{figure}

In Fig.~\ref{fig:ts_k} we compare the model spectra of Toor \& Seward with the
combined nebula $+$ pulsar spectrum of K01. It is seen that in the
MECS band (1--10~keV) both models agree within 2~\%; the differences are
higher near 100~keV (up to 10~\%) and at low energies (8~\% at 0.1~keV).

\subsection{Intrinsic curvature of the Crab nebula spectrum\label{sect:curcor}}

The K01 power-law model for the nebula that we use here is reasonably well
established for energies above 2~keV. However, below 2~keV interstellar
absorption becomes important. The usual approach with low-resolution instruments
has been to assume that the power-law extends unbroken down to about 0.5~keV,
and to adjust the interstellar column density until a good spectral fit at low
energies is obtained. This approach was also used by K01.

There is no guarantee, however, that the power-law is unbroken below 2~keV. The
evidence that the spectrum is not a simple power-law comes from both theoretical
considerations and observations.

\begin{figure}
\resizebox{\hsize}{!}{\includegraphics[angle=-90]{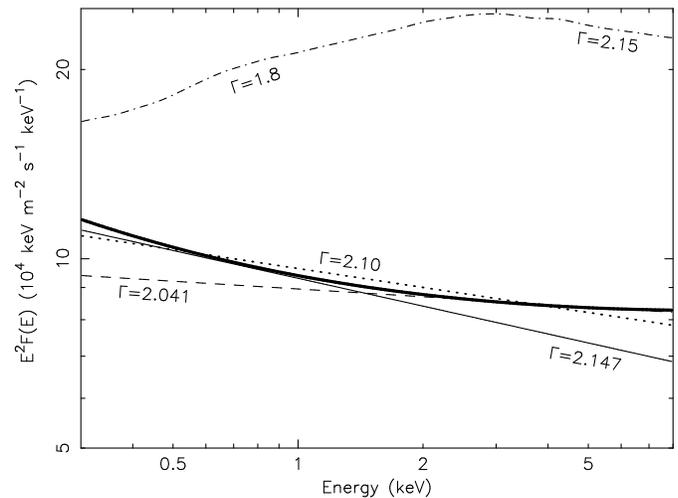}}
\caption{Unabsorbed continuum spectrum of the Crab nebula. 
Thin solid line ($\Gamma=2.147$): power-law model of \citet{kuiper2001}. 
Thick solid line: integrated spectrum derived from the Chandra maps of photon
index and count rate as given by \citet{mori2004}. 
Dashed line ($\Gamma=2.041$): power-law approximation to the above
integrated Chandra spectrum in the 2--8~keV band. 
Dotted line ($\Gamma=2.10$): direct power-law fit of \citet{mori2004} to
the full nebula spectrum. 
Dash-dotted line: theoretical model spectrum (run A) determined by
\citet{delzanna2006}. }
\label{fig:cont_comp}
\end{figure}

\subsubsection{Theoretical considerations}

The Crab nebula emits synchrotron emission due to the motion of accelerated
particles in a magnetic field. By its nature, such radiation has a high-energy
cut-off corresponding to the maximum energy of the particles. 
\citet{delzanna2006} have made detailed magneto-hydrodynamical models for pulsar
nebulae with synchrotron emission incorporated self-consistently. Their model A
was chosen to have parameters relevant for the Crab nebula. That model describes
the spectral morphology of the nebula qualitatively well. The integrated
spectrum shows multiple breaks due to spatially varying  maximum particle
energies. In the X-ray band, their integrated model spectrum has a break around
3~keV, with a slope of 1.8 below it and 2.15 above it (dash-dotted line in
Fig.~\ref{fig:cont_comp}). However, this model over-predicts the observed X-ray
flux at 1~keV by a factor of 2, and another model B -- although less consistent
with the morphology of the Crab nebula -- has spectral breaks at different
energies. What we gather from these models is therefore that spectral breaks in
the integrated X-ray spectrum may occur between 0.5 and 10 keV, but the details
of where they occur and how strong they are are hard to predict exactly.

\subsubsection{Observational evidence of spectral curvature\label{sect:obscurv}}

Evidence of spectral breaks also comes from spectral imaging studies of the Crab
nebula. There are clear spatial variations of the photon index over the surface
of the nebula, as for example observed with XMM-Newton \citep{willingale2001}.
Using Chandra ACIS-S data, \citet[M04 hereafter]{mori2004} fitted simple
power-laws to the local spectrum of small regions and produced a map of the
total count rate and photon index $\Gamma$. They fixed the Galactic absorption
to $N_{\mathrm H} = 3.2\times 10^{25}$~m$^{-2}$. The photon index ranges between
1.5 and 3.1, implying that the integrated spectrum must show curvature. We have
converted the 0.5--8~keV count rate map of M04 into flux using version 3.7 of
PIMMS for calculating the photon-index dependent conversion curve. The total
nebular spectrum was calculated by adding the individual spectra of the small
regions. This total spectrum indeed shows curvature (thick solid line in
Fig.~\ref{fig:cont_comp}), although over the 0.3--8~keV band it never deviates
more than 5~\% from the simple power-law with $\Gamma=2.10$ as determined
directly from a spectral fit to the integrated spectrum (M04).

Our aim is to determine how much the low-energy continuum is enhanced relative
to the high-energy power-law spectrum. We saw that the 1--100~keV integrated
nebula spectrum as determined by other instruments shows no breaks (K01;
Sect.~\ref{sect:nebulamodel}). K01 put some question marks to the robustness of
their fit below 1~keV, because of the complexity of the spectral/spatial
redistribution function of the BeppoSAX LECS in that band. Moreover, below 2~keV
the interstellar absorption affects the spectrum more than 2~\% (below 1~keV
more than 8~\%) introducing additional uncertainties.  Therefore we determine
the spectral curvature relative to the well-determined 2--8~keV continuum.

\begin{figure}
\resizebox{\hsize}{!}{\includegraphics[angle=-90]{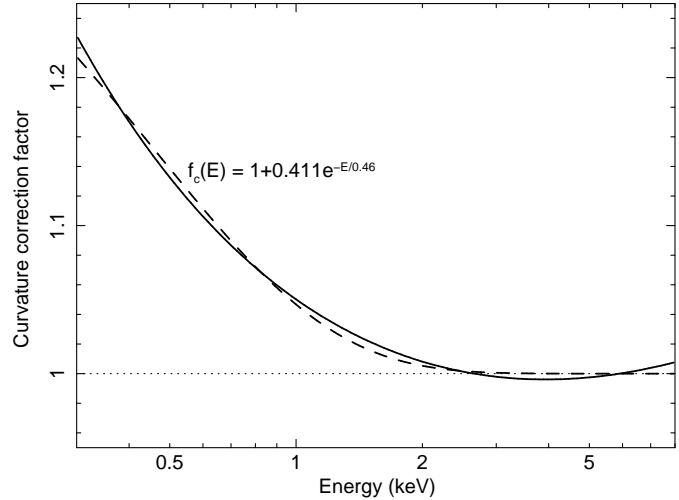}}
\caption{Solid line: correction factor $f_c(E)$ for the intrinsic curvature of
the  continuum spectrum of the Crab nebula, obtained by comparing the full 
integrated model spectrum to a power-law approximation in the 2--8~keV band.
Dashed line: analytical approximation to $f_c(E)$  (Eqn.~\ref{eqn:fc}).}
\label{fig:fcurcor}
\end{figure}

The spatially integrated curved spectrum in the 2--8~keV band is approximated by
a single power-law with $\Gamma=2.041$, to within 1~\% accuracy (see
Fig.~\ref{fig:cont_comp}). We define the intrinsic curvature correction function
$f_c(E)$ as the ratio of the full integrated spectrum to this 2--8~keV power-law
approximation. The correction factor $f_c(E)$ is shown in
Fig.~\ref{fig:fcurcor}. At $E=0.5$~keV, the integrated spectrum is 13~\% higher
than the power-law approximation, showing the importance of the correction. We
will use a simple analytical approximation for $f_c(E)$: 
\begin{equation} 
f_c(E) = 1 + 0.411 {\mathrm e}^{\displaystyle{-E/0.46}} 
\label{eqn:fc} 
\end{equation} 
where $E$ is the photon energy in keV. The accuracy of this approximation is
better than 1~\%, and over the RGS band (5--38~\AA) it is even better than
0.6~\%.

The curvature correction derived above (\ref{eqn:fc}) represents the steepening
of the soft X-rays with respect to the hard X-ray power-law spectrum, for the
full Crab spectrum. The photon index of the nebula at high energies
($2-8$~keV) based on the Chandra data (2.041) differs by 0.106 from the photon
index that we obtained in Sect.~\ref{sect:nebulamodel} based on K01 (2.147, see
also Fig.~\ref{fig:cont_comp}). This difference may look surprisingly high, but
can be explained well. The Chandra data were fitted in a low energy band (below
8~keV), where Galactic absorption cannot be neglected; moreover, the Galactic
column density obtained by M04 is $4.1\times 10^{24}$~m$^{-2}$ lower than the
value used by K01. This leads to different ISM transmission factors that can be
approximated to lowest order by a power-law over the $0.3-8$~keV band with a
slope of 0.087. Thus if M04 would have used a higher Galactic column density
(like the K01 value), their photon indices would have been higher by 0.087,
close to and within the systematic uncertainties of the K01 value. The question
then arises why M04 arrived at a lower column density. This may be due to the
spectral softening of the intrinsic nebula continuum associated with the
curvature correction; furthermore, M04 derived their column density from the
fainter regions of Crab, which are generally in the outer parts with
intrinsically steeper spectra.

As the curvature correction gives the {\sl relative change} in spectral index
between the soft and hard band, and is based on the spatial variations of the
local photon index, an overall change of all local photon indices due to
accounting for systematic uncertainties does, to first order, only affect
the integrated hard and soft band spectral slopes, but not their difference. In
a similar way, a different treatment of the interstellar absorption will affect
all spectral slopes in the same way and will not affect the curvature
correction. Therefore, we apply the curvature correction (\ref{eqn:fc}) also to
the K01 continuum.

As an independent test, we have used a MOS data set to derive the curvature
correction in a similar way as described above for Chandra (we use the MOS
observation of Appendix~\ref{sect:dispcor}). Within the statistical errors, we
obtain the same curvature correction. See Appendix~\ref{sect:curcorunc} for more
details and estimates of the uncertainties on the curvature correction.

The total, unabsorbed continuum spectrum of Crab (nebula plus pulsar)
is then given by
\begin{equation}
F_p(E) + F_n(E) f_c(E),
\end{equation}
where $F_p(E)$, $F_n(E)$ and $f_c(E)$ are given by (\ref{eqn:pulsar}),
(\ref{eqn:nebula}) and (\ref{eqn:fc}), respectively. The systematic uncertainty
on the slope of Crab at high energies ($F_p(E) + F_n(E)$) is estimated in
Sect.~\ref{sect:sys_crab} to be 0.016. The systematic uncertainties on the
curvature correction $f_c(E)$ are estimated in Sect.~\ref{sect:curcorunc}, and
give rise to systematic slope uncertainties lower than 0.010 over the RGS
band.

In order to show the effects of this curvature correction, in our analysis of
the RGS spectrum we also show fit results for the case of the  K01 unbroken
power-law without the curvature correction given by (\ref{eqn:fc}). 

\section{Observations and data analysis\label{sect:data}\label{sect:analysis}}

\subsection{Data selection}

\begin{table}
\begin{center}
\caption{Observation log.}
 \begin{tabular}[t]{cccccc}
 \hline
 \hline
ID$^a$ & seq. & RGS    & CCDs   & $t_{\mathrm{exp}}$ & $\Delta\theta$$^b$ \\
     &   nr   & 1 or 2 & in use & (ks)               & (arcmin) \\
 \hline
 101 & 002  & 1 & 1       & 15.5 & - \\
 101 & 003  & 1 & 2       & 15.2 & - \\
 101 & 004  & 1 & 3       &  7.0 & - \\
 101 & 005  & 1 & 4       &  3.7 & - \\
 101 & 006  & 1 & 5       &  1.6 & - \\
 101 & 007  & 1 & 6       &  0.6 & - \\
 101 & 008  & 1 & 8       &  0.5 & - \\
 101 & 009  & 1 & 9       &  1.5 & - \\
 101 & 033  & 2 & 1       & 15.5 & - \\
 101 & 034  & 2 & 2       & 15.2 & - \\
 101 & 035  & 2 & 3       &  7.0 & - \\
 101 & 036  & 2 & 5       &  3.7 & - \\
 101 & 037  & 2 & 6       &  1.6 & - \\
 101 & 038  & 2 & 7       &  0.6 & - \\
 101 & 039  & 2 & 8       &  0.5 & - \\
 101 & 040  & 2 & 9       &  1.5 & - \\
 101 & 010  & 1 & 1,5     &  4.0 & - \\
 101 & 011  & 1 & 2,6     &  4.5 & - \\
 101 & 012  & 1 & 3,8     &  4.0 & - \\
 101 & 013  & 1 & 4,9     &  4.5 & - \\
 101 & 041  & 2 & 1,6     &  4.0 & - \\
 101 & 042  & 2 & 2,7     &  4.5 & - \\
 101 & 043  & 2 & 3,8     &  4.0 & - \\
 101 & 044  & 2 & 5,9     &  4.5 & - \\
 101 & 014  & 1 & 1,3,5,9 &  3.9 & - \\
 101 & 015  & 1 & 2,4,6,8 &  2.5 & - \\
 101 & 045  & 2 & 1,3,5,7 &  2.5 & - \\
 101 & 046  & 2 & 2,6,8,9 &  2.9 & - \\
 101 & 016  & 1 & all     &  1.7 & - \\
 101 & 047  & 2 & all     &  1.9 & - \\
 201 & 004  & 1 & all     &  5.8 & $+3.5$ \\
 201 & 005  & 2 & all     &  5.8 & $+3.5$ \\
 301 & 004  & 1 & all     &  5.7 & $-3.5$ \\
 301 & 005  & 2 & all     &  5.7 & $-3.5$ \\
 401 & 004  & 1 & all     &  8.5 & $+7.0$ \\
 401 & 005  & 2 & all     &  8.5 & $+7.0$ \\
 501 & 004  & 1 & all     & 14.9 & $-7.0$ \\
 501 & 005  & 2 & all     & 14.9 & $-7.0$ \\
\hline
\end{tabular}
\label{tab:obslog}
\end{center}
$^a$ Observation ID (the leading digits 0312790 are the same for all
observations and have been omitted here)\\
$^b$ off-axis angle in the cross-dispersion direction
\end{table}

The data were taken on February 24 and 25, 2006, during orbit 1138 of
XMM-Newton. We only discuss here the RGS observations. For each observation,
either 1, 2, 4 or 8 CCDs were read out in a fixed sequence. This setup was done
in order to study the effects of pile-up on the spectrum (see
Appendix~\ref{sect:pileup}). In addition, a set of observations with off-axis
pointing in the cross-dispersion direction was done, with off-axis angles of
$\pm 3.5$ and $\pm 7.0$~arcmin, in order to study the spatial distribution of
the emission in the cross-dispersion direction (see
Appendix~\ref{sect:crosscor}). A summary of the observations is shown in
Table~\ref{tab:obslog}. The observations were processed using SAS version 7.0
and the set of calibration files (CCF) available October 2006. More information
on the spectral extraction and the various correction factors that we need are
given in Appendix~A.

\subsection{Spectral fitting\label{sect:specmodel}}

All spectral fitting was done using the SPEX package \citep[see also
www.sron.nl/spex]{kmn}. RGS1 and RGS2 data were fit simultaneously. The
continuum model for Crab was taken to be the K01 model (pulsar plus nebula)
discussed in Sect.~\ref{sect:pulsarmodel} and \ref{sect:nebulamodel}, including
the curvature correction derived in Sect.~\ref{sect:obscurv}. This continuum
model was read as a table (model {\sl file} in SPEX), and the normalisation of
this continuum was frozen. We apply two multiplicative components to this
continuum, discussed below: a correction for the unknown effective area of the
RGS, and a model for the interstellar absorption. Therefore in our fit, the only
free parameters are the properties of the interstellar absorption (column
densities and line width), and the two parameters of the effective area
correction.

As we indicated in Appendix~\ref{sect:chebychev}, the effective area of RGS has
been determined modulo a power-law correction. The true effective area
$A_{\mathrm{true}}(\lambda)$ as a function of the wavelength $\lambda$ (in \AA)
can be related to the effective area currently present in the response matrix
$A_{\mathrm{resp}}(\lambda)$ by $A_{\mathrm{true}}(\lambda)=f(\lambda)
A_{\mathrm{resp}}(\lambda)$ where \begin{equation} f(\lambda) =
A(\lambda/10)^\alpha. \label{eqn:aeff_cor} \end{equation} For technical reasons
we have chosen not to fit directly the parameters $A$ and $\alpha$, but to fit
use $f(\lambda_1)$ and $f(\lambda_2)$ with $\lambda_{1,2}$ extreme points for
the RGS band, chosen here as $5$ and $40$~\AA.

For our absorption model, we use a combination of the {\sl hot} and {\sl slab}
models of SPEX. Both models use the same set of atomic absorption cross sections
(see below). The {\sl slab} model calculates the transmission of a layer of
plasma with arbitrary composition. Free parameters are the column densities of
individual ions as well as the intrinsic velocity broadening. We use this model
component for all neutral atoms of the elements that show clearly detectable
absorption edges in the RGS band (N, O, Ne, Mg and Fe), plus their singly
ionised ions (for detections or upper limits to the ionised fraction). 

For all other elements we use the {\sl hot} model. This model calculates the
transmission of a plasma in collisional ionisation equilibrium. We use this with
a very low temperature (the neutral gas limit). For the abundances of those
elements we adopt the proto-solar abundances of \citet{lodders2003}. The
dominant contributors in the RGS band for those elements are He, H, C, Si and S
(in order of importance). The cross sections of all these elements are tied to
the hydrogen cross section.

The cross sections used for both the {\sl hot} and {\sl slab} models are
described in more detail in the SPEX manual. Basically, both continuum and line
opacity are taken into account. For neutral atoms in the RGS band predominantly
inner shell absorption lines are important; the cross sections of those line
were mainly calculated by E. Behar using the HULLAC code; see also Section 5.1
of \citet{kaastra2002} for more details.

Alternatively, in order to investigate the role of dust, we have also made
spectral fits with the model outlined above (which we label model A), but with a
part of the total column replaced by a gas and dust mixture (model B). This gas
and dust mixture is assumed to have also the proto-solar abundances of
\citet{lodders2003}, but a fraction of the gas is depleted into dust grains. The
depletion factors of the dust are taken from \citet{wilms2000}. The transmission
of the dust is calculated using the {\sl dabs} model that we introduce here in
the SPEX package, which follows completely the dust treatment as described by
\citet{wilms2000} and references therein. Basically, the grains are assumed to
be spherical, fluffy ($\rho = 1000$~kg\,m$^{-3}$), and have a size distribution
${\mathrm d}n/{\mathrm d}a \sim a^{-p}$ for the grain radius $a$ between
$a_{\min} < a < a_{\max}$, with default values $a_{\min} = 0.025~\mu$m,
$a_{\max} = 0.25~\mu$m, and $p=3.5$.

As a third model (Model C) we also made a fit without the curvature correction
but further identical to model A, in order to investigate the effects of this
curvature correction.

For the spectral fitting, we rebinned all data by a factor of 5 to account for
the reduced spectral resolution caused by the spatial extent of Crab
(0.22~\AA\ FWHM, see Appendix~\ref{sect:dispcor}). This gives bins with a width
of $\sim 0.04$~\AA\ at 7~\AA\ to $0.07$~\AA\ at 30~\AA. Fitting was restricted
in most cases to the 7--30~\AA\ band; below 7~\AA,  scattering by the RGS
gratings becomes very important, but theory of scattering at these short
wavelengths starts to break down. Therefore, there is essentially no knowledge
of the RGS effective area below 7 \AA. Hence this wavelength range was excluded
from the fit. Above 30~\AA, the low energy of the X-ray events deposited on the
CCD, combined with the extent of the source, starts to coincide with the tail of
the CCD system peak. For this reason, PI selection was limited at low energies,
as explained in Appendix~\ref{sect:extraction}. Still, some contamination of
X-ray events by the CCD system peak can be expected. For this reason, the
spectrum above 30~\AA\ was also excluded from the spectral fit (see
Fig.~\ref{fig:best1}). However, we were able to derive the nitrogen abundance by
extending the fit range -- for the sole purpose of determining that abundance --
to 32~\AA. That fit gives biased values for all other parameters, but the jump
across the nitrogen edge at 30.8~\AA\ is determined well.

Furthermore, we omitted small regions near the interstellar oxygen K-edge
(22.7--23.2~\AA) and iron L-edge (17.2--17.7~\AA), for reasons explained in more
detail in our discussion (Sect.~\ref{sect:feedge}--\ref{sect:oedge}; essentially
the presence of molecules that affect the fine structure of the edge). As the
ISM abundances are mainly determined from the depth of the absorption edges,
ignoring these small regions does not affect the basic outcome of our analysis.

\subsection{Results of the spectral fits\label{sect:specfit}}.

\begin{figure*}[!htb]
\begin{center}
\resizebox{\hsize}{!}{\includegraphics[angle=-90]{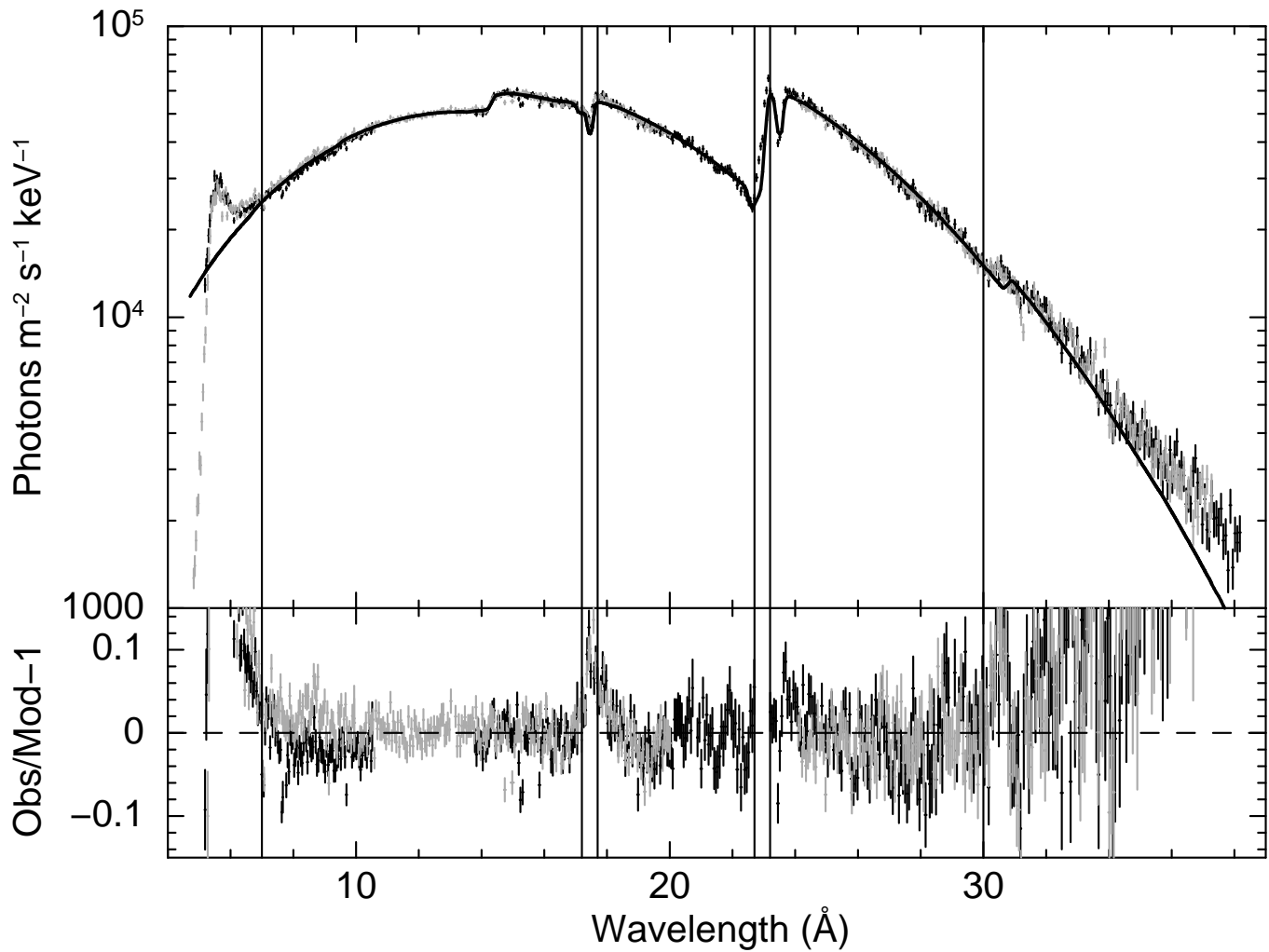}}
\caption{Best fit RGS spectrum of the Crab nebula using model A. The fit was
restricted to the 7--30~\AA\ range (indicated by the vertical lines), excluding
two small regions near the \ion{O}{i} K-edge and \ion{Fe}{i} L-edge (also
indicated by vertical lines). However in the plot we show the full data range.
Single and double CCD readout modes have been combined, but RGS1 (dark colour)
and RGS2 (light colour) data are shown separately. The upper panel shows the
fluxed RGS spectrum (but at the original spectral resolution of the data,
including spectral broadening due to the spatial extent of the nebula). For
clarity reasons we plot the flux in keV units along the y-axis, which reduces
the dynamical range along that axis significantly. The lower panel shows the
relative fit residuals.}
\label{fig:best1}
\end{center}
\end{figure*}

We have obtained a fit using model~A outlined in Sect.~\ref{sect:specmodel} with
a $\chi^2$ of 2341 for 1316 degrees of freedom. The fit is shown in
Fig.~\ref{fig:best1} and the parameters are given in Table~\ref{tab:fitparam}.
The fit residuals over the fitted range have -- on average -- an intrinsic
systematic scatter of 1.8~\% and a random noise component of 1.9~\%, yielding a
total scatter of 2.6~\%. Therefore, the typical systematic uncertainty in our
model is less than 2~\%. The higher deviations outside the fitted range can be
simply understood as we already discussed in Sect.~\ref{sect:specmodel}.

\begin{table}
\begin{center}

\caption{Best fit parameters. }

 \begin{tabular}[t]{lccc}
 \hline
 \hline
 parameter$^a$ & model A     & model B     & model C \\
               & pure gas    & gas \& dust & pure gas \\
	       & curv. corr. &  curv. corr &  no curv. corr \\    
 \hline
 $\chi^2$      & 2341 & 2327 & 2366 \\         
 d.o.f.        & 1316 & 1315 & 1316 \\        
 $f(5)^{b}$    & $1.056\pm 0.009$ & $1.055\pm 0.009$ & $1.035\pm 0.009$ \\
 $f(40)^{b}$   & $0.746\pm 0.016$ & $0.762\pm 0.016$ & $0.854\pm 0.019$ \\
 \hline
\multicolumn{4}{l}{column densities$^{c}$:}\\  	 	 
 \hline
\ion{H}{i}$^{d,e}$ & $31750\pm 220$ & $ 9700 \pm 1700 $ & $31280\pm 220$ \\
\ion{H}{i}$^{d,f}$ &        $-$     & $22300 \pm 1800 $ & $-$  \\
\ion{H}{i}$^{d,g}$ & $31750\pm 220$ & $32060\pm 220   $ & $31280\pm 220$ \\
\ion{H}{ii}$^h$ &$1752\pm 4$& $1752\pm 4$ & $1752\pm 4$\\
\ion{N}{i}     & $ 2.67\pm 0.24$ & $ 2.72\pm 0.24$ & $ 2.48\pm 0.25$ \\
\ion{O}{i}     & $19.28\pm 0.18$ & $19.83\pm 0.22$ & $19.23\pm 0.18$ \\
\ion{O}{ii}    & $ 0.23\pm 0.14$ & $ 0.25\pm 0.15$ & $ 0.20\pm 0.14$ \\
\ion{Ne}{i}    & $ 4.69\pm 0.21$ & $ 4.66\pm 0.21$ & $ 4.74\pm 0.21$ \\
\ion{Ne}{ii}   & $ 0.46\pm 0.25$ & $ 0.52\pm 0.24$ & $ 0.64\pm 0.25$ \\
\ion{Mg}{i}    & $ 1.14\pm 0.28$ & $ 1.20\pm 0.28$ & $ 1.13\pm 0.28$ \\
\ion{Mg}{ii}   & $ 0.00\pm 0.09$ & $ 0.00\pm 0.10$ & $ 0.00\pm 0.09$ \\
\ion{Fe}{i}    & $ 0.65\pm 0.05$ & $ 0.77\pm 0.04$ & $ 0.65\pm 0.05$ \\
\ion{Fe}{ii}   & $ 0.23\pm 0.06$ & $ 0.14\pm 0.04$ & $ 0.26\pm 0.07$ \\
\hline
\end{tabular}
\label{tab:fitparam}
\end{center}
$^a$ Error bars are all $1\sigma$ for one interesting parameter.\\
$^b$ $f(5)$ and $f(40)$ are the values of the effective area scale factor 
$f(\lambda)$, Eqn.~(\ref{eqn:aeff_cor}), at 5 and 40~\AA.\\ 
$^c$ Ionic column densities are given in units of $10^{21}$~m$^{-2}$. The column
densities of all metals include all phases (dust and gas).\\ 
$^d$ \ion{H}{i} includes in the
cross section contributions from neutral He, C, Si, S, Ar, Ca and Ni assuming
proto-solar abundances for these elements.\\
$^e$ For the pure gaseous phase only.\\
$^f$ For the gas \& dust mixture with proto-solar abundances added
to the pure gaseous phase above.\\
$^g$ Total column: sum of contributions listed as $b$ and $c$.\\
$^h$ For \ion{H}{ii} the column is derived from the pulsar dispersion measure
assuming that all free electrons originate from hydrogen (see text).\\
\end{table}

For our alternative model B that includes dust, we list the best fit parameters
also in Table~\ref{tab:fitparam}. We have tried to leave the basic dust
parameters free in our fits: grain mass density $\rho$, minimum and maximum size
$a_{\min}$ and $a_{\max}$, and the size distribution index $p$. However, our fit
is not very sensitive to these parameters, and therefore we have kept them to
their default values listed in Sect.~\ref{sect:specmodel}.

\subsubsection{Search for hot gas\label{sect:hotgassearch}}

The fit residuals near 21.6~\AA\ (Fig.~\ref{fig:on_edge}) and 19.0~\AA\
(Fig.~\ref{fig:nefe_edge}), where we would expect the 1s--2p transitions of
\ion{O}{vii} and \ion{O}{viii} respectively, show some evidence of additional
weak absorption features. We have determined the formal equivalent width of
these features by adding narrow absorption lines to the model (but broadened due
to the spatial extent of Crab as seen by RGS). 

For the \ion{O}{vii} line, we find a formal equivalent width of $9\pm 4$~m\AA,
and for \ion{O}{viii} $7\pm 2$~m\AA\ if the line is narrow. For \ion{O}{viii}
the feature looks somewhat broader; if we allow the width of the line to vary,
we find a higher equivalent width of $18\pm 4$~m\AA, but this requires a high
velocity width (Gaussian $\sigma = 5000$~km\,s$^{-1}$).

For consistency, we have checked a Chandra LETGS spectrum of the Crab pulsar
(Obsid 759), and found a formal equivalent width of the \ion{O}{vii} line of
$7\pm 17$~m\AA, consistent with our RGS measurement.

\section{Discussion of the absorption spectrum of Crab\label{sect:discussion}}

\subsection{Fine structure near absorption edges}

Our fits (Fig.~\ref{fig:best1}) clearly show strong absorption edges of neutral
O, Fe and Ne. Other edges are weaker (like Mg at 9.47~\AA\ and N at 30.77~\AA).
In principle, the neutral Si edge at 6.72~\AA\ is also in the RGS band and is
expected to have an optical depth of 0.017; however, it is in the part of the
spectrum where our fits become poor and also any uncertainty related to the
instrumental Si edge will contribute to its systematic uncertainty. Therefore we
will not discuss Si further in our paper. We will discuss here in more detail
the edges of Ne, Fe, and O.

\subsubsection{Neon K-edge}

\begin{figure}[!htb]
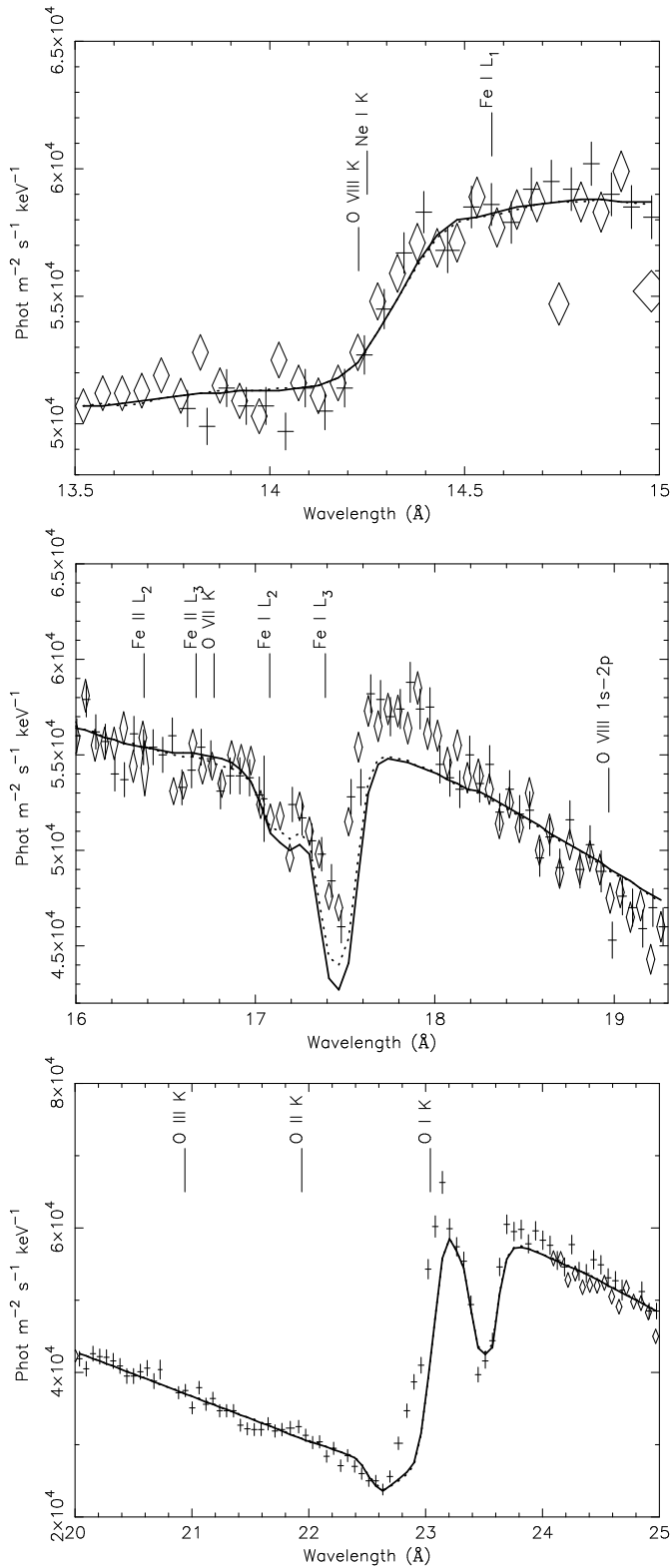

\begin{center}
\resizebox{\hsize}{!}{\includegraphics[angle=-90]{ne_edge.ps}}
\resizebox{\hsize}{!}{\includegraphics[angle=-90]{fe_edge.ps}}
\resizebox{\hsize}{!}{\includegraphics[angle=-90]{o_edge.ps}}
\caption{Blow-up of Fig.~\ref{fig:best1} near the \ion{Ne}{i} K-edge (top),
\ion{Fe}{i} L$_{2,3}$-edge (middle) and \ion{O}{i} K-edge (bottom).
Crosses: RGS1; diamonds: RGS2; solid line: model A (no dust); dotted line:
model B (with dust). }
\label{fig:nefe_edge}
\label{fig:on_edge}
\end{center}
\end{figure}  
 
The Ne-edge is clearly detected in the data (Fig.~\ref{fig:nefe_edge}). The
optical depth $\tau$ of the edge is 0.15. Nearby is the L$_1$ edge of neutral
iron, but that edge is much weaker: $\tau=0.010$. Recently, \citet{juett2006}
have studied the fine structure near the neon K-edge and iron L-edge using
Chandra HETGS spectra of X-ray binaries. The spectral resolution of the HETGS
(0.01--0.02~\AA), combined with the point source nature of these objects, gives
them effectively a 10--20 times higher spectral resolution than our Crab
spectrum (extended source). For \ion{Ne}{i}, in addition to the pure absorption
edge, there is the series of 1s--$n$p transitions ($n>2$ because the 2p shell is
completely filled for \ion{Ne}{i}). According to the model of Juett et al., the
1s--3p line is at $14.295\pm 0.003$~\AA, and this line is strong enough that the
effective K-edge of neon shifts towards this wavelength. For the true
edge\footnote{The definition of "edges" by Juett et al. is non-standard.
They define the edge where empirically the flux drops; we keep here the more
formal definition of the series limit of the absorption lines.}, we used the
value of 14.249~\AA. When we leave it as a free parameter, the best fit value
for the edge is $14.231\pm 0.013$~\AA. This low shift is most likely caused by
our neglect of the Rydberg series for $n>5$ in our absorption model.

The K-edges of \ion{O}{viii} and \ion{Ne}{i} differ only by $\sim 0.02$~\AA.
With our current effective spectral resolution for Crab, we cannot resolve these
edges. However, as we show  later (Sect.~\ref{sect:hotgasdiscussion}) the amount
of hot gas in the line of sight is small enough to be neglected.

\subsubsection{Iron L-edge\label{sect:feedge}}

We have already seen that the L$_1$ edge of \ion{Fe}{i} is very weak, and we
will not discuss that edge further. The predicted depth of the L$_2$ and $L_3$
edges for our best-fit model A are only $\tau=0.027$ and $\tau=0.056$,
respectively. At a first glance this looks contradictory to the deep features
seen in Fig.~\ref{fig:nefe_edge}. The difference is due to the presence of
strong inner-shell line transitions (for example 2p--3d excitations). Such
transitions have first been identified in an astrophysical source by 
\citet{sako2001}, and have subsequently been incorporated in our absorption
models in SPEX. Those transitions cause the two deep absorption troughs at 17.1
and 17.4~\AA. They are also visible in the HETGS data of \citet{juett2006}, at
similar wavelengths. At our effective spectral resolution it is not possible to
separate the inner shell transitions of \ion{Fe}{i} from those of \ion{Fe}{ii};
the blends for these ions almost coincide. Therefore, the difference between the
absorption by both ions is mainly derived from the different location of the
absorption edges (see Fig.~\ref{fig:nefe_edge}).

Similar to the case of the neon edge, also here possible contamination by highly
ionised oxygen must be considered. Within our effective resolution, the
\ion{O}{vii} K-edge is close to the \ion{Fe}{ii} $L_3$ edge
(Fig.~\ref{fig:nefe_edge}). Our model gives an optical depth of 0.019 for this
\ion{Fe}{ii} $L_3$ edge, while the \ion{O}{vii} edge has a depth of at most
0.002. Thus, contamination of the iron edge by highly ionised oxygen is less
than 10~\%, well below the statistical uncertainty on the \ion{Fe}{ii} column
density.

Although the column density of Fe is determined well from the jump across the 
absorption edge, our fit is not optimal. At the long wavelength side of the
edge, between 17.5--18.0~\AA, there is excess flux of the order of 5~\%. This
excess extends downwards to 17.1~\AA. Adopting a $\sim 5$~\% higher continuum
between 17--18~\AA, our absorption model provides a reasonable fit. Note that
our model B with dust provides a slightly better fit in the absorption lines
compared to model A; this is due to the high opacity of the lines, which results
into stronger shielding effects by dust grains, and therefore to a relatively
lower line to continuum opacity than for the pure atomic case.

The 5~\% amplitude of the excess is much higher than any uncertainties in our
cross-dispersion correction factor, excluding therefore an origin in
uncertainties of dust scattering near the iron edge. As Crab has a rather smooth
synchrotron continuum, we do not expect such broad features to be present in the
intrinsic Crab spectrum. 

It is striking, however, that several of the fits of \citet{juett2006} to
spectra of X-ray binaries have excess flux of a similar magnitude just at the
long wavelength side of the Fe edge, similar to our case (for example
\object{Cyg X-1}). Although their features look a little sharper, convolving
them with the spatial/spectral response of our Crab observation would yield
similar features.

The instrumental fluorine K-edge near 17.8~\AA\ is determined well from
higher resolution RGS spectra of point sources with low interstellar absorption
columns. Those spectra do not show similar residuals. Hence we can exclude an
origin of the excess in the instrumental F K-edge.

Finally we note that it is unlikely that this excess is due to line emission
from a supernova shell. \citet{seward2006}, using Chandra data, have derived an
upper limit to the luminosity of such a shell of $2\times 10^{27}$~W. The
observed excess with RGS, when fitted with a Gaussian profile, has a flux of
$109\pm 8$~ph\,m$^{-2}$\,s$^{-1}$, at a mean wavelength of $17.53\pm 0.03$~\AA\
and with $\sigma=0.57\pm 0.08$~\AA. At the distance of Crab, and correcting for
foreground absorption, this corresponds to a (line) luminosity of $2.4\times
10^{27}$~W, almost equal to the upper limit to the total X-ray luminosity of any
shell. Moreover, RGS cannot see all flux from such a shell unless the radius is
less than 2.5\arcmin. For such a radius, the maximum wavelength shift observed
with RGS due to the spatial extent of the emission is 0.35~\AA. There are no
strong line transitions close enough to explain the emission. For example, if
the emission would be \ion{O}{viii} Ly$\alpha$ emission at 18.97~\AA, it should
be located about 10\arcmin\ from the Crab pulsar.

\subsubsection{Oxygen K-edge\label{sect:oedge}}

The oxygen K-edge clearly is the strongest feature of the spectrum, with a jump
of a factor of 2 across the edge. The 1s--2p absorption line is reproduced well
by our fit. However, in our original fit the edge energy, for which we used a
value of 23.045~\AA, clearly seems to be off. Making it a free parameter, we
find a best-fit wavelength of $22.917\pm 0.008$~\AA, a shift of
$-0.128$~\AA. This does not mean that the edge itself is shifted. Two effects
may cause the discrepancy. 

First, we use the continuum cross sections of \citet{verner1995}. In this
calculation, there is no fine structure at the edge, which is located at 538~eV
(23.045~\AA). The edge, however, has at least two strong sub-edges at the $^2P$
series limit (22.567~\AA), and $^4P$ series limit (22.767~\AA).  Furthermore,
the calculated wavelengths of the absorption lines that we use cannot be
calculated more accurately than typically 10--30~m\AA. Only the strongest
transition (1s--2p) was adjusted empirically to observed values. All this causes
significant uncertainty in the adopted absorption model for atomic oxygen. We do
not attempt to resolve this problem here but refer to a forthcoming paper
\citep{devries2008} for a more extensive discussion of the K-shell absorption
structure in X-ray sources; also see the discussion in \citet{paerels2001} and
\citet{juett2004}.

Secondly, we note that in our fit with model B, with our adopted oxygen dust
depletion factor of 0.4 for the dust$+$gas mixture, the total fraction of oxygen
bound in dust is 28~\% (as there is also a pure gaseous component). This number 
is quite uncertain as it depends on our adopted dust depletion factor. Oxygen
bound in minerals or dust grains has a clearly different fine structure near the
edge than atomic oxygen; see for instance the many examples in
\citet{vanaken1998}. This changes the absorption structure for the chemically
bound oxygen population significantly.

As the \ion{O}{ii} and \ion{O}{iii} edges are separated well from the 
\ion{O}{i} edge, the high spectral resolution of RGS allows us to put tight
upper limits on the amount of ionised oxygen. No strong edges of these ions are
seen in the spectrum.

Fig.~\ref{fig:on_edge} also shows a small depression near 21.6~\AA, the
wavelength of the strong resonance line of \ion{O}{vii}. We discuss that further
in Sect.~\ref{sect:hotgasdiscussion}.

Finally, based on our work on blazars as described in paper III
\citep{paper3}, we estimate that the systematic uncertainty on the neutral
oxygen column in Crab due to the uncertainty in the instrumental oxygen edge
is less than 1~\%.

\subsection{Column densities of the cold gas\label{sect:columns}}

From the column densities obtained from Table~\ref{tab:fitparam} we see that the
absorbing medium is mostly neutral, except for iron, for which we find a
significant ionised contribution. Column densities towards Crab have also been 
measured using UV and optical spectroscopy by \citet[S00
hereafter]{sollerman2000}. They derived column densities based on a Doppler
width  of 1~km\,s$^{-1}$. For such a low velocity broadening, the lines are
strongly saturated and hence the statistical uncertainty on the column densities
are relatively high, despite significant line detections.

S00 obtained a neutral hydrogen column density of $(3.0\pm 0.5) \times
10^{25}$~m$^{-2}$, fully consistent with our more precise value of
$(3.18\pm 0.02) \times 10^{25}$~m$^{-2}$ for model A or $(3.21\pm 0.02) \times
10^{25}$~m$^{-2}$ for model B. 

There are 4 ions in common between our work and S00: \ion{O}{i}, \ion{Mg}{i},
\ion{Mg}{ii} and \ion{Fe}{ii}.  For \ion{O}{i} and \ion{Fe}{ii}, the column
densities agree within the statistical errors of a factor of 2 and 4,
respectively.  However, for \ion{Mg}{i} and \ion{Mg}{ii} there is a bigger
difference. S00 find a \ion{Mg}{ii}/\ion{Mg}{i} ratio of 5, with an uncertainty
of a factor of 2, while we find a ratio less than 0.1. The total magnesium
columns however agree: we find $(1.1-1.2\pm 0.3)\times 10^{25}$~m$^{-2}$ for
models A and B respectively, compared to $(0.7^{+0.7}_{-0.4}) \times
10^{25}$~m$^{-2}$ for S00.

The different ratio appears puzzling but can be understood as follows. First,
the \ion{Mg}{i} and \ion{Mg}{ii} edges are close, at 9.48 and 9.30~\AA. The
optical depth of the \ion{Mg}{i} edge (our deepest Mg edge) is about 0.02. In
this region of the spectrum (see Fig.~\ref{fig:best1}), there is a small
difference between the RGS1 and RGS2 data, at the 1~\% level. It is therefore
easy to get systematic errors at this level. While the total magnesium column is
rather robust (as it measures essentially the jump across the combined nearby
\ion{Mg}{i} and \ion{Mg}{ii} edges), discriminating between both ions is harder
because this depends solely on the small data stretch between 9.30--9.48~\AA. 

We therefore believe that a higher \ion{Mg}{ii}/\ion{Mg}{i} ratio than reported
by us is quite well possible, but that our total magnesium column is rather
secure. Such a higher ratio would also be consistent with the enhanced degree of
ionisation of iron.

\subsection{ISM constituents}

For a proper discussion of abundances, the different phases of the ISM should be
taken into account. Near the Sun, we have the following components (see the
recent overview of the ISM by \citet{ferriere2001} for more details):
\begin{enumerate}
\item hot ionised gas ($\sim 10^6$~K),
\item warm ionised gas ($\sim 8000$~K),
\item warm atomic gas (6\,000--10\,000~K),
\item cold atomic gas (20--50~K),
\item molecular gas (10--20~K),
\item and dust
\end{enumerate}
We consider these components in more detail below.

\subsubsection{Hot ionised gas\label{sect:hotgasdiscussion}}

In general, the presence of hot gas in the line of sight towards any X-ray
source is most easily seen using high spectral resolution measurements of the
associated absorption lines. This is because in general the column densities of
highly ionised gas are relatively low and therefore the absorption edges are
weak, while the absorption lines still may be visible.

X-ray absorption due to the presence of hot gas in our Galaxy has been reported
before. \citet{yao2006} analysed Chandra grating spectra toward the bright low
mass X-ray binary \object{4U~1820-303}, which has a neutral hydrogen column
density of $1.5\times 10^{25}$~m$^{-2}$, half of the value towards Crab. In
4U~1820-303, these authors find equivalent widths for the \ion{O}{vii}
and \ion{O}{viii} 1s--2p absorption lines of $40\pm 12$ and  $19\pm 6$~m\AA,
respectively. These values are much higher than the values that we obtain for
Crab (Sect.~\ref{sect:hotgassearch}), of $9\pm 4$ and $7\pm 2$~m\AA,
respectively.

Due to the weakness of the \ion{O}{vii} and \ion{O}{viii} absorption lines
toward Crab, it is hard to estimate the associated column densities. If the
velocity dispersion of the gas $\sigma_{\mathrm v}=180$~km\,s$^{-1}$, the value
found by Yao \& Wang toward 4U~1820-303, then the column densities of
\ion{O}{vii} and \ion{O}{viii} are $(8\pm 3) \times 10^{19}$ and $(2.6\pm 0.9)
\times 10^{19}$~m$^{-2}$, respectively, 2.5 and 10 times below the values
towards 4U~1820-303. The optical depth at the K-edge of these ions is then less
than 0.002, and therefore the neglect of these ions in our analysis is
justified. For solar abundances, and assuming that all oxygen of the hot gas
would be in the ionisation stage of \ion{O}{vii} or \ion{O}{viii}, the total
hydrogen column density of the hot gas would be $2\times 10^{23}$~m$^{-2}$, less
than 1~\% of the neutral hydrogen column density.

Unfortunately, we do not have a good constraint on $\sigma_{\mathrm v}$. If 
$\sigma_{\mathrm v}$ were substantially lower than 180~km\,s$^{-1}$, much
higher columns would be required. We have worked this out as follows. We made a
fit to our spectrum adding columns of \ion{O}{vii}, \ion{O}{viii} and
\ion{Ne}{ix} to our model in the form of an additional {\sl slab} component.
These ions are dominant X-ray opacity sources for typical temperatures of a
million K. Surprisingly, our fit improves significantly ($\Delta\chi^2=88$). The
improvement is produced mainly by the inclusion of the absorption lines as well
as fine tuning of the continuum, in particular near the neon and iron edges,
although the $\sim 5$~\% excess emission near the Fe-L edge still remains.
Because the \ion{O}{vii} K-edge is close to the \ion{Fe}{ii} L$_{2,3}$ edges
(see Fig.~\ref{fig:nefe_edge}), the fit reduces the \ion{Fe}{ii} column to zero
in exchange to the \ion{O}{vii} column. In a similar way, \ion{Ne}{i} is reduced
in favour of \ion{O}{viii}, also because both edges are so close
(Fig.~\ref{fig:nefe_edge}). Still the overabundance of neon remains.

However, this model is not acceptable because of other reasons. First, the total
column of highly ionised oxygen (\ion{O}{vii} and \ion{O}{viii}) is high:
$5\times 10^{21}$~m$^{-2}$. Combined with the ionised hydrogen column deduced
from the electron density (determined from the pulsar dispersion measure, see
Sect.~\ref{sect:warmgas}), an overabundance of oxygen of 6 times solar is
required. This overabundance is even a lower limit if a significant fraction of
the ionised hydrogen is associated with warm ($10^4$~K like) gas instead of hot
($10^{6}$~K like) gas. The most important problem with this model is, however,
that the required velocity dispersion of the gas is $\sigma_{\mathrm
v}<3$~km\,s$^{-1}$, far below the thermal velocity of the hot gas. If the hot
gas is predominantly in collisional ionisation equilibrium, then for a typical
temperature of $10^6$~K, the thermal velocity of oxygen ions is more than
20~km\,s$^{-1}$. 

Of course, we can solve this problem by accounting properly for the thermal
width of the lines. We have done this by replacing the additional {\sl slab}
component by a {\sl hot} component with solar abundances. Additional free
parameters are the total hydrogen column density, the temperature and the
velocity broadening $\sigma_{\mathrm v}$ to be added in quadrature to the
thermal width of the lines. This model yields only a modest fit improvement over
the original model without hot gas: $\Delta\chi^2 = 23$. We find a typical
temperature of 2--3 million K and a hydrogen column of $5\times
10^{23}$~m$^{-2}$, almost two orders of magnitude below the column density of
neutral hydrogen. Accordingly, the optical depths of the strongest oxygen edges
for this component are at most 0.001, and there is no significant change to any
of the parameters for the cold absorption component.

In summary, we conclude that some hot gas may be present in the line of sight
towards Crab, at a level of a percent of the neutral hydrogen column. Typically,
the amount of hot gas towards Crab is less than $N_{\mathrm H} \la 5\times
10^{23}$~m$^{-2}$. This is small enough that neglecting it does not affect our
results for the cold phase or the calibration of the Crab spectrum.

As the abundances in the hot ionised gas may be slightly different from those of
the colder components \citep{yao2006}, we exclude this hot component explicitly
from our abundance analysis.

\subsubsection{Warm ionised gas\label{sect:warmgas}}

Pulsar dispersion measurements \citep[for example][]{lundgren1995} yield a total
electron column density of $(1.752\pm 0.003)\times 10^{24}$~m$^{-2}$, where the
error is dominated by the monthly fluctuations of this number, probably caused
by the Crab nebula itself \citep{isaacman1977}. As this electron column density
is high compared to the hot ionised gas column density, most of the free
electrons must be produced by the warm ionised gas. For the typical conditions
prevailing in that medium, most of these electrons come from hydrogen. Hence, we
use the electron column density as an estimate for the \ion{H}{ii} column
density.  

The other elements in the warm ionised gas are either neutral or singly ionised,
and they are accounted for in our absorption model, although we cannot
distinguish between the neutral atoms in this gas phase and those in the
cold and warm atomic gas.

\subsubsection{Cold and warm atomic gas}

These two gas phases mainly contain neutral atoms. In the X-ray band, they are
hard to distinguish, the main difference being the thermal line width. These are
hard to measure, however, as the associated absorption lines are strongly
saturated. Therefore we discuss the combined abundance of both phases, and also
include the neutral and singly ionised atoms from the warm ionised gas in this
component.

\subsubsection{Molecular gas}

Special attention needs to be paid to molecular hydrogen (H$_2$) here, as it is
the main molecular X-ray opacity source.

The distribution of molecular gas in the Galaxy is very irregular and depends
strongly upon the line of sight. It has a relatively high concentration closer
to the centre of the Galaxy and is less important further outwards. In the local
ISM, about 20--25~\% of the hydrogen is in its molecular form
\citep{savage1977}, and it is this value that is used as default in for example
the absorption model of \citet{wilms2000}, although these authors warn the user
to take care of such spatial variations. Over the X-ray band, the H$_2$ opacity
per hydrogen atom is about $\sigma_{\mathrm{mol}}/\sigma_{\mathrm{atom}} = 1.42$
times higher than for \ion{H}{i} \citep{wilms2000}. Hence, for a mixture with a
20~\% molecular contribution, the effective hydrogen opacity should be 8~\%
higher than for pure \ion{H}{i}; ignoring this molecular component would
therefore bias abundance determinations.

In our spectral model we have ignored molecules, so we need to justify this.
We do this by determining $R$, the relative enhancement of the total opacity
including molecules compared to the opacity of the same gas if it is fully
atomic. $R$ can be written as
\begin{equation}
R = (\frac{\sigma_{\mathrm{mol}}}{\sigma_{\mathrm{atom}}} - 1)
 \times 
\frac{n_{\mathrm{mol}}}{n_{\mathrm{atom}}}
 \times 
f_{\mathrm H}
\end{equation}
where here $n_{\mathrm{mol}}=2 n_{\mathrm{H}2}$ and $n_{\mathrm{atom}}$ are the
number densities of hydrogen nuclei in molecular and atomic form, respectively.
Furthermore, $f_{\mathrm H}$ is defined here as the effective opacity of hydrogen
relative to the other metals in our X-ray spectrum.

For gas with solar composition, the contribution of hydrogen to the total
opacity varies between 5~\% at 10~\AA\ to 17~\% at 30~\AA. A typical weighted
average value for our Crab spectrum is $\sim 10$~\%. As approximately half of
the total opacity is produced by elements such as N, O, Ne, Mg and Fe for which
we determine the column densities separately, we estimate $f_{\mathrm H}$ by
comparing the pure hydrogen opacity to the total opacity produced by the
elements that we tied to hydrogen using solar abundances. This leads to an
estimated of $f_{\mathrm H}=0.2$.

Crab lies in the direction of the Galactic anticentre, at a distance of 2~kpc.
Therefore the fraction of H$_2$ is expected to be lower than the typical value
of 20~\%. From the CO maps of \citet{dame2001} (available through NASA's skyview
facility), and using their conversion factor from CO equivalent width to H$_2$
column, we find $n_{\mathrm{H}2} <5\times 10^{22}$~m$^{-2}$. There is a
molecular cloud with a higher column density of $10^{24}$~m$^{-2}$ North of
Crab, but it is more than 5\arcmin\ away. Therefore, our best estimate is
$n_{\mathrm{mol}}/n_{\mathrm{atom}} <0.003$ (direct upper limit) or
$n_{\mathrm{mol}}/n_{\mathrm{atom}} <0.06$ (worst case scenario, if the cloud
North of Crab would cover it).

Taking all these factors into account, we find $R<0.42\times 0.003 \times 0.2 
<3\times 10^{-4}$. For the worst case scenario with
$n_{\mathrm{mol}}/n_{\mathrm{atom}} <0.06$, still $R<0.005$. Therefore, we can
completely ignore the role of molecular gas in our line of sight.

\subsubsection{Dust}

Dust in the line of sight towards Crab or any other X-ray source has two
effects: scattering and absorption of X-rays. 

Scattering does not destroy photons but merely redistributes them into a halo
around the source. Recently, \citet{seward2006} have studied the dust scattering
halo of Crab using Chandra imaging. They conclude that the total column density
of scattering material (dust and associated gas) is $1.7-2.2\times
10^{25}$~m$^{-2}$ for their best fit model, and 2/3 of that value for an
alternative model. Most of this can be modelled to be in a smoothly distributed
component between us and Crab, while 10--25~\% of the scattering material is in
a single structure at a distance of $\sim 100$~pc from the Sun, perhaps
associated with the boundary of the local hot bubble. Surprisingly, the total
column density that \citet{seward2006} find is less than the column derived from
X-ray absorption measurements. This may indicate a relative low abundance of
dust towards Crab.

The amount of dust inside Crab is relatively small, and can contribute only
1--2~\% to the total column density towards Crab. This can be derived from
infrared observations: \citet{temim2006} estimate that Crab contains about
0.004--0.010~M$_{\odot}$ of graphite or 0.006--0.015~M$_{\odot}$ of silicates,
in order to explain the thermal dust emission seen at 70~$\mu$m. Assuming a
constant density sphere with radius 2\arcmin, the total carbon column of the
dust within Crab (in case of graphite) would be $0.7-1.8\times
10^{20}$~m$^{-2}$, and in case of silicates (assuming for simplicity olivine or
Fe$_2$SiO$_4$) the silicon column would be $0.6-1.6\times 10^{19}$~m$^{-2}$. In
both cases, this corresponds to 1--2~\% of the total interstellar column. Hence
we will ignore this component.

Apart from scattering, dust also absorbs X-rays. There are two main differences
between the absorption properties of atoms contained in dust and free atoms.
First, the absorption edges and lines in bound atoms are slightly different, due
to the chemical shifts. Secondly, the effective absorption cross section can be
less due to shielding in grains.

Unfortunately, our spectral resolution is not very high due to the extended
nature of the Crab emission. Therefore it is hard to measure accurately the
location of absorption lines. However, our data still contain some usefull
information on that. 

\begin{table}
\begin{center}
\caption{Wavelengths of absorption lines, all in \AA.}
\begin{tabular}[t]{@{}l@{\,}c@{\quad}c@{\quad}c@{\quad}c@{}}
 \hline
 \hline
 Constituent                     & oxygen   & iron & iron & ref. \\
                                 & main line & main line & 2nd line & \\
 \hline
 Ferrous, Fe$^{2+}$              &         & 17.498 & 17.196 & a \\
 Olivine                         & 23.26   &        &        & b \\
\hline
 Ferric, Fe$^{3+}$               &          & 17.456 & 17.130 & a \\
 Hematite                        & 23.43    &        &        & b \\
 \hline
 Atomic \ion{O}{i}               & 23.508   &        &        &  c \\ 
 Atomic \ion{Fe}{i}              &          & 17.453 & 17.142 &  d \\ 
\hline
 Crab    & $23.466\pm 0.009$ & $17.396\pm 0.009$ & $17.120\pm 0.016$ & e \\
\hline
\end{tabular}
\label{tab:abslines}
\end{center}
$^a$ \citet{vanaken2002}, corrected by $-0.019$~\AA\ following \citet{juett2006} \\
$^b$ \citet{vanaken1998} \\
$^c$ \citet{juett2004}, from Chandra spectra \\
$^d$ \citet{behar2001}, HULLAC calculations \\
$^e$ Present work
\end{table}

In Table~\ref{tab:abslines} we compile our current knowledge about the dominant
absorption lines in different constituents, and compare it with the observed
centroids as measured in our Crab spectrum. These observed wavelengths were
obtained by ignoring the absorption lines in our best fit model and fitting them
instead with narrow absorption lines. Olivine (Mg$_{2-x}$Fe$_x$SiO$_4$) is one
of the best known examples of a ferrous, and hematite (Fe$_2$O$_3$) of a ferric
compound.

If we consider neutral oxygen, then the observed absorption line centroid
suggests approximately equal contributions from atomic and ferric iron
(Table~\ref{tab:abslines}). Such a mixture also gives good agreement with the
secondary iron line blend at 17.12~\AA, and is also in agreement with the $\sim
30$~\% of oxygen bound in dust derived from our model B. There is only
disagreement with the main iron line blend at 17.40~\AA. Clearly, both ferric
and atomic iron have their predicted line closer to the observed centroid of
this blend than ferrous iron, but none of them agree fully. It may be that the
predicted wavelengths by \citet{behar2001} are slightly off, as these lack
detailed experimental validation; however, their calculations agree within a few
m\AA\ with other calculations (Raassen \& Kaastra, private communication, as
quoted by Behar et al.). We suggest that the dust line of sight towards Crab is
dominated by ferric compounds.

\begin{figure}[!htb]
\begin{center}
\resizebox{\hsize}{!}{\includegraphics[angle=-90]{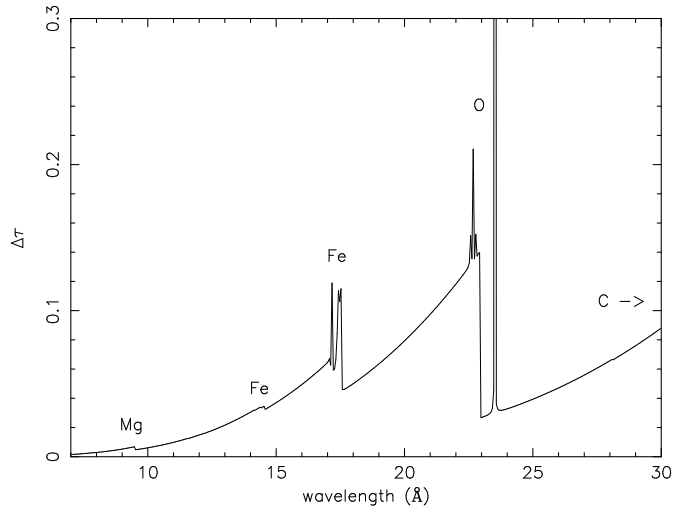}}
\caption{Difference in optical depth for an equivalent hydrogen column density
of $3.16\times 10^{25}$~m$^{-2}$, for dust with depletion factors as given by
\citet{wilms2000}, compared to the same material but dissolved into free atoms.
The most important spectral features are indicated. }
\label{fig:diftau}
\end{center}
\end{figure} 

The second difference in absorption properties of dusty atoms compared to free
atoms is the shielding of atoms in dust grains. The shielding leads to
effectively lower opacity per atom. This is illustrated in
Fig.~\ref{fig:diftau}. At the oxygen edge, the difference can be up to 14~\%. In
practice, the effects of inclusion of dust in the model are not as dramatic as
shown here. The reason is that one can always adjust the abundances slightly in
order to correct for the highest changes in the depth of the edges. Therefore,
despite the significance of the different dust cross sections, the total opacity
is similar to that of pure atomic gas and this is the main reason of the only
modest improvement of the fit with model B compared to model A. It also explains
why it is hard to use our data to constrain the properties of the dust --
density, grain size, etc.

However, the most robust statement we can make about the dust is the associated
column density. From Table~\ref{tab:fitparam} we see that the column density of
the gas associated with the dust is about 70~\% of the total column density, and
its value of $(2.23\pm 0.18) \times 10^{25}$~m$^{-2}$ is in good agreement with
the value derived by \citet{seward2006} based on modelling of the dust
scattering halo. Thus, we confirm their finding of a reduced dust column towards
Crab as compared to a standard gas/dust mixture.

\subsection{ISM abundances}

\begin{table}
\begin{center}
\caption{Abundances, all relative to the proto-solar values
of \citet{lodders2003}. }
 \begin{tabular}[t]{lccc}
 \hline
 \hline
 Element & Model A      & Model B     & Model C\\
         & pure gas     & gas \& dust & pure gas \\
	 & curv. corr.  &  curv. corr &  no curv. corr \\    
\hline
H    & $\equiv 1$      & $\equiv 1$       & $\equiv 1$       \\
N    & $1.00\pm0.09$   & $1.01\pm 0.09$   & $0.95\pm 0.10$   \\
O    & $1.012\pm0.014$ & $1.030\pm 0.016$ & $1.022\pm 0.014$ \\
Ne   & $1.72\pm 0.11$  & $1.72\pm 0.11$   & $1.83\pm 0.11$   \\ 
Mg   & $0.82\pm 0.21$  & $0.85\pm 0.21$   & $0.82\pm 0.21$   \\
Fe   & $0.76\pm 0.07$  & $0.78\pm 0.05$   & $0.79\pm 0.07$   \\
\hline
\end{tabular}
\label{tab:abun}
\end{center}
\end{table}

We have determined the abundances of all elements for which we have reliable
individual column density measurements. These abundances are shown in
Table~\ref{tab:abun}. They include all phases of the ISM except for the hot
ionised gas. We see that the abundances do not depend significantly on the
inclusion of dust in our models. 

Interestingly, the abundances that we derive for most elements except for neon
are close to the proto-solar abundances of \citet{lodders2003} that we used as a
reference. Oxygen and nitrogen fully agree with the proto-solar abundances,
while the magnesium and iron abundances are only slightly lower. 

The only exception to this general abundance pattern is neon. It has an
overabundance of $\sim 1.7$. \citet{antia2005} and \citet{drake2005} recently
argued that the solar neon abundance should be higher. Based on measurements of
neon to oxygen in stars, Drake \& Testa derive a Ne/O number ratio of 0.41. 
Relative to the proto-solar values (Ne/O number ratio 0.15), this would
imply an overabundance of neon compared to oxygen of 2.6.  Clearly, we find an
enhanced neon abundance, but only 2/3 of the value of Drake \& Testa. In
fact, our Ne/O number ratio of $0.26\pm 0.02$ is in excellent agreement with
recent measurements of the abundances in the Orion nebula (Ne/O$=0.24\pm 0.05$,
\citealt{esteban2004}) and of B-stars in that nebula (Ne/O$=0.26\pm 0.07$,
\citealt{cunha2006}).

\subsubsection{The role of carbon\label{sect:carbon}}

As the carbon K-edge is longwards of the RGS wavelength band, it is difficult to
estimate the carbon abundance from our measurements. However, as carbon
contributes approximately 10~\% to the total opacity below the oxygen K-edge, it
is potentially an important factor to take into account. 

Our best-fit model predicts a total carbon column density of $9.2\times
10^{21}$~m$^{-2}$ for our adopted proto-solar abundances. S00 found from their
optical and UV spectroscopy a column density of atomic \ion{C}{i} of $8.5\times
10^{21}$~m$^{-2}$; however the error bars of a factor of 2 allow a range between
$4-16\times 10^{21}$~m$^{-2}$. In addition, they found a \ion{C}{ii} column of
$14\times 10^{21}$~m$^{-2}$, but here the error bars are a factor of 3, allowing
a range between $5-50\times 10^{21}$~m$^{-2}$. Given these numbers and the fact
that a part of the carbon may be contained in dust, it is not very likely that
the total column density of carbon is much below the proto-solar value. 

To estimate the possible effects of a different carbon abundance, we have
determined the best fit parameters in case the carbon abundance decreases to 0.7
times proto-solar, a similar value as the iron and magnesium abundance towards
Crab. $\chi^2$ of our fit then decreases by only 1.4, the hydrogen column
density increases by $10^{24}$~m$^{-2}$ or 3.2~\%,  $f(5)$ increases by 0.3~\%
and $f(40)$ decreases by 1.7~\%. We will use these numbers in our estimate of
the systematic uncertainty in the RGS effective area.

\section{Discussion of the effective area calibration of the RGS\label{sect:rgs}}

We use our spectral fit to determine the absolute effective area of the RGS.
This is achieved by determining the power-law correction factor $f(\lambda)$
from Eqn.~(\ref{eqn:aeff_cor}). As we described in Sect.~\ref{sect:specmodel},
we determine this from the values of $f(\lambda)$ at 5 and 40~\AA. From
Table~\ref{tab:fitparam} we see that we can determine these numbers to an
accuracy of 0.9 and 1.6~\%, respectively. The differences of these values for
model A and B are not very high, at most 2~\% at the longest wavelengths.

The differences between model A and B with model C are much higher, as model C
ignores the intrinsic curvature of the underlying continuum. But despite that
the curvature correction at 40~\AA\ (0.3~keV) is 22~\% (see
Fig.~\ref{fig:fcurcor}), the enhancement of f(40) is only 12~\%. This is because
also other parameters in the fit, in particular the neutral hydrogen column,
have changed. The lower hydrogen column for model C partially compensates for
the neglect of the continuum enhancement at low energies. Comparing model C with
model A, none of the metallic column densities neither the abundances differ by
more than 1$\sigma$. However, as we have shown in Sect.~\ref{sect:curcor}, there
are strong arguments for applying the curvature correction. The systematic
uncertainties on the correction as derived in Appendix~B are significantly lower
than the correction factor itself. We therefore propose to adopt our model B as
the model that agrees best with all constraints.

The relatively low statistical uncertainties on $f(\lambda)$ are not the
most important errors, however. There is a range of systematic factors that
increase the uncertainty on $f(\lambda)$. The most obvious one is of course the
10~\% uncertainty on the absolute flux of the K01 intrinsic continuum that we
used (Sect.~\ref{sect:intrinsic}). In another paper \citep{paper3} we show how
we can reduce that uncertainty by combining the present results from our Crab
analysis with observations of white dwarfs, for which we can determine the
absolute X-ray flux with an accuracy of a few percent. Other systematic
uncertainties cannot be reduced, however. In order to be able to assess the
accuracy of our calibration, we need to make a quantitative assessment of all
these systematic effects. 

We estimate the known systematic uncertainties and present them in
Table~\ref{tab:syserr}. We proceed as follows. For each relevant factor, we 
estimate its magnitude for a range of characteristic wavelengths, covering the
RGS range between 5--30~\AA. The systematic uncertainties are calculated by
comparing the nominal fluxes with the derived fluxes when one of the relevant
parameters is changed as described in Appendix~B. These numbers are given in
the table. Then we put each error in one of two specific categories. 

For category "c", the systematic uncertainties at different wavelengths are more
or less correlated. Example: the absolute flux of Crab is known to only 10~\%.
Therefore, if the true flux of Crab at 5~\AA\ differs by 10~\% from the value we
adopted, it will differ by the same percentage for all other wavelengths. Taking
this into account, we can assess for these "correlated" errors how they will
affect the normalisation and slope of $f(\lambda) = A(\lambda/10)^\alpha$ by
"fitting" the estimated systematic uncertainties (all having the same sign)
directly to a power-law shape. 

For category "u", the systematic uncertainties at different wavelengths are
uncorrelated over the full RGS wavelength scale. A good example are the
cross-dispersion corrections: as we use a combination of splines and power-law
approximations, systematic errors are correlated over ranges between 0.5 and
several \AA, but there is no relation between the uncertainty at 7~\AA\ to that
at 30~\AA. In those cases, we estimated $\Delta A$ from the typical  slope and
normalisation that we would obtain if the systematic uncertainties at different
wavelengths had random signs. In general, for this type of uncertainty the
direct effect on $A$ and $\alpha$ is smaller than for correlated errors
(category "c").

More details on the individual systematic uncertainties are given in Appendix~B.

\begin{table*}
\begin{center}
\caption{Summary of systematic and statistical uncertainties in the Crab spectrum. }
 \begin{tabular}[t]{rc|c|cccc|ccc|cccc|c}
 \hline
 \hline
 $\lambda$ & $E$ &   a   &   b   &   c   &   d   &   e   &   f   &   g   &   h   &   i   &   j   &   k   &   l    &  m \\
 \AA\      & keV &       &       &       &       &       &       &       &       &       &       &       &        &    \\
\hline
 5 & 2.5         & 0.1   & 0.008 & 0.003 & 0.000 & 0.000 & 0.001 & 0.001 & 0.003 & 0.050 & 0.000 & 0.30  & 1.0    & 0.0031 \\
 7 & 1.8         & 0.1   & 0.011 & 0.005 & 0.000 & 0.003 & 0.002 & 0.002 & 0.000 & 0.010 & 0.001 & 0.04  & 0.022  & 0.0025 \\
10 & 1.2         & 0.1   & 0.015 & 0.007 & 0.001 & 0.005 & 0.004 & 0.002 & 0.004 & 0.004 & 0.002 & 0.01  & 0.010  & 0.0025 \\
15 & 0.8         & 0.1   & 0.019 & 0.010 & 0.001 & 0.000 & 0.006 & 0.010 & 0.008 & 0.005 & 0.002 & 0.00  & 0.008  & 0.0026 \\
20 & 0.6         & 0.1   & 0.022 & 0.011 & 0.002 & 0.005 & 0.007 & 0.001 & 0.010 & 0.015 & 0.001 & 0.00  & 0.008  & 0.0045 \\
25 & 0.5         & 0.1   & 0.024 & 0.013 & 0.002 & 0.005 & 0.009 & 0.003 & 0.012 & 0.020 & 0.001 & 0.00  & 0.008  & 0.0048 \\
30 & 0.4         & 0.1   & 0.026 & 0.014 & 0.003 & 0.002 & 0.010 & 0.001 & 0.014 & 0.050 & 0.000 & 0.00  & 0.010  & 0.0090 \\
\hline
\multicolumn{2}{c|}{type}
                 &   c   &   c   &   c   &   c   &   u   &   c   &   c   &   c   &   u   &   c   &   u   &   u    & u \\
\multicolumn{2}{c|}
{$\Delta\alpha$} & 0.016 & 0.010 & 0.006 & 0.002 & 0.001 & 0.005 & 0.001 & 0.010 & 0.010 & 0.000 & 0.027 & 0.006  & - \\
\multicolumn{2}{c|}
{$\Delta A$}     & 0.100 & 0.019 & 0.010 & 0.001 & 0.001 & 0.004 & 0.005 & 0.004 & 0.003 & 0.002 & 0.005 & 0.002  & - \\
\end{tabular} 
\label{tab:syserr}
\end{center}
{\sl Main power-law:} \\
$^a$ Uncertainty in the K01 continuum model \\
{\sl Curvature corrections:} \\
$^b$ A 50~\% change in the lower energy of the 2--8~keV range
used for the normalisation of the Chandra data \\
$^c$ Break size $\Delta\Gamma=0.006$ at low $E$ 
if a break (at 4~keV) is present in the Kuiper et al. data \\
$^d$ 0.5~\% of the absorption with $N_{\mathrm H}$ 
higher by $7.5\times 10^{24}$~m$^{-2}$ \\
$^e$ Numerical approximation of $f_c(E)$ by (\ref{eqn:fc}) \\
{\sl Interstellar absorption:} \\
$^f$ Systematic uncertainty in the dust model \\
$^g$ Neglect of hot gas \\
$^h$ Interstellar carbon abundance \\
{\sl Instrumental corrections:} \\
$^i$ Cross-dispersion correction \\
$^j$ Pile-up correction for 2 CCD mode; adopts 5~\% uncertainty in $a$ ($2\sigma$) \\
$^k$ Redistribution errors, estimated from difference fit residuals RGS1 and RGS2 \\
$^l$ Errors in the Chebychev polynomials used for the effective area correction \\
$^m$ Typical combined RGS1/RGS2 statistical uncertainty for a full 1~\AA\ wide bin \\
\end{table*}

For completeness, we also list in Table~\ref{tab:syserr} the statistical
uncertainties on the Crab spectrum. As can be seen, these are low compared to
the systematic uncertainties.

The main systematic uncertainty in the slope of our power-law correction ($\pm
0.027$) originates from the redistribution errors (column (k) of
Table~\ref{tab:syserr}). These uncertainties are mostly important below 10~\AA.
In principle, it should be possible to reduce these errors further, by a
carefull analysis of a large sample of sources with different intrinsic spectral
shape in the hard band. However, such an effort is beyond the scope of this
paper. If we add up all other systematic uncertainties $\Delta\alpha$ on the
slope in quadrature, except for the redistribution errors, the combined
uncertainty is 0.020; the total r.m.s. systematic error is 0.034.

The main uncertainty on the absolute flux is -- apart from the uncertainty in
the K01 model listed in column (a) -- the uncertainty in the lower energy used
for the scaling of the Chandra spectra on the K01 model (column (b).

In another paper \citep{paper3} we show how this uncertainty in flux can be
reduced to about 3~\% by combining our results for Crab with white dwarf
spectra. However, none of that analysis affects our present results on the Crab
nebula.

\section{Conclusions\label{sect:conclusions}}

We have studied the interstellar absorption towards Crab and determined the
effective area of the RGS. 

The interstellar line of sight towards Crab appears to be free of molecules and
has a relatively low contribution of hot gas. It contains only about 70~\% of
the amount of dust expected from its absorption column, a conclusion that is in
agreement with studies of the dust scattering halo by  \citet{seward2006}. From
the centroids of the dominant O-K and Fe-L absorption features, we conclude that
gas and dust contribute roughly equally to these features, and that ferric iron
is more abundant in the dust than ferrous iron.

We have determined the interstellar abundances with very high precision, the
oxygen abundance with a relative accuracy even better than 2~\%. While oxygen
and nitrogen are fully compatible with proto-solar abundances, we find slightly
lower ($\sim 80$~\% proto-solar) abundances for iron and magnesium, although for
the last element we cannot exclude proto-solar abundances. On the other hand,
neon is over-abundant by a factor of 1.7. 

We can also put tight limits on the amount of singly ionised ions, thanks to our
high spectral resolution. For instance, less than 2~\% of all oxygen
is in the form of \ion{O}{ii}.

Concerning the calibration of RGS, we have accurately determined the effective
area of this instrument. The relative accuracy of the effective area over the
7--30~\AA\ band is equivalent to a power-law index uncertainty of about 0.03.
The absolute effective area from our analysis is known to the same accuracy as
the adopted continuum model spectrum of Crab, namely 10~\%. In a forthcoming
paper \citep{paper3} we will show how that last uncertainty can be reduced to a
few percent.

\begin{acknowledgements}

We thank Andy Pollock and his colleagues at ESOC for helpful comments on this
paper. This work is based on observations obtained with XMM-Newton, an ESA
science mission with instruments and contributions directly funded by ESA Member
States and the USA (NASA). SRON is supported financially by NWO, the Netherlands
Organization for Scientific Research. 

\end{acknowledgements} 
 
\bibliographystyle{aa}
\bibliography{crab}

\begin{appendix}

\section{Spectral extraction and response generation}

\subsection{Spectral extraction\label{sect:extraction}}

Spectra and response matrices were created using the standard XMM-Newton data
analysis system (SAS), version 7. The standard RGS pipeline 'rgsproc' was used,
but with a few adaptations explained below.

For point sources, the preferred method is to extract spectra that contain 95~\%
of all photons in the cross-dispersion direction, and 95~\% of all photons in
the pulseheight-wavelength ({\sl PI}, "banana") regions. The response matrix
then corrects for the missing flux fraction in both subspaces.

In our case, the situation is different. Crab is an extended source, with
substantial emission wings containing $\sim$10~\% of all flux in the ROSAT band.
These wings are visible up to half a degree from Crab \citep{predehl1995}. Crab
is therefore larger than the width of the RGS detector in the cross-dispersion
direction, and also photons are spread over a larger region in the dispersion
direction than for a point source. Moreover, Crab is bright and for some parts
of the source pile-up of events cannot be neglected.

For these reasons, we extracted source spectra over the full width of the RGS
detector (5.0 arcmin), and with a {\sl PI} selection that contains 98~\% of the
flux for each wavelength. Selecting 100~\% {\sl PI} is not possible because then
spectral order separation becomes impossible and at low {\sl PI} values the
selected range would fall below the lower threshold.

To prevent the broader {\sl PI} selection to select background events from the
CCD system peak, the {\sl PI} selection was limited to {\sl PI} values higher
than 200 eV, well above the average system peak of around 100 eV. This made the
{\sl PI} selection narrower than 98~\% for wavelengths longer than 24.8~\AA.
This limitation of PI selection for a particular wavelength range is properly
taken into account by the SAS processing system. 

The response matrices that we used, however, were created for a selection of
95~\% in {\sl PI} space and 95~\% in cross-dispersion space. These are the
recommended values for point sources. We use those matrices because they are the
best calibrated (see Sect.~\ref{sect:chebychev}), but we apply correction
factors in the form of an {\sl arf} file to this standard response matrix. Note
that in general the standard SAS software produces only {\sl rmf} matrices where
the effective area has already been included. As our corrections go beyond the
standard SAS, we have chosen here for the {\sl arf} option.

The correction factors entering the {\sl arf} are discussed in the next
sections. All factors are multiplied to get the total area correction. They are:

\begin{enumerate}
\item $0.98/0.95$, to correct for the 95~\% {\sl PI} selection in the
standard response matrix as compared to our actual 98~\% extraction.
\item $1.00/0.95$, to correct for the 95~\% cross-dispersion selection
in the standard response matrix as compared to our full detector (100~\%)
extraction.
\item Dead time can be ignored. See Sect.~\ref{sect:deadtime}.
\item Pile-up correction $f_p(\lambda)$. See Sect.~\ref{sect:pileup}.
\item Vignetting of telescope and gratings. See Sect.~\ref{sect:vignetting}.
\item Loss of photons in the cross-dispersion direction due to the extended
size of Crab $f_{\mathrm d}(\lambda)$. See Sect.~\ref{sect:crosscor}.
\end{enumerate}

Finally, we need to correct for the extent of the source in the dispersion
direction. This causes some spectral resolution degradation, but does
not affect the effective area. This is discussed in Sect.~\ref{sect:dispcor}.

\subsection{Standard response matrix\label{sect:chebychev}}

The standard response matrix contains an empirical correction factor for the
nominal effective area for point sources (95~\% {\sl PI} selection, 95~\%
cross-dispersion selection). This empirical correction is based on all RGS
spectra from \object{Mrk~421}, used as calibration source. We assume Mrk~421 to
have a pure power-law spectrum  of arbitrary power-law index and normalisation,
which may change over time. We fit a power-law in the 10-25~\AA\ wavelength
range. The remaining part of the RGS spectrum is made to fit this spectrum using
a set of low frequency Chebychev correction polynomials (up to order 12). A
time-dependant correction is added assuming an increasing Carbon deposit on the
RGS CCD's, with parameters based on the constant sources
\object{RX~J1856.5$-$3754} and the \object{Vela pulsar}. 

The standard SAS calibration now contains the Chebychev correction  polynomials
and carbon absorption model including the absolute effective area calibration
based on the work described in this article. This latter correction was of
course not used for this analysis, since this is a factor to be determined.  

\subsection{Dead time effects\label{sect:deadtime}}

For sources producing high count rates in the RGS such as the Crab nebula, many
events are lost due to onboard processing constraints. When too many events are
queued for on board processing or telemetry download, the processor will discard
a number of CCD readouts (frames) to free its buffer. This will influence dead
time, but will have no other consequences for the instrument response. The data
reduction SAS software will take the exact number of lost frames into account to
compute the proper exposure time. Net result will only be a higher statistical
noise than is expected on the basis of observing time only.

Since we selected 100~\% of the CCD width, events recorded during CCD frame
transfer (out-of-time events) would still fall within the spatial selection
window and hence do not contribute to any dead time.

\subsection{Pile-up corrections\label{sect:pileup}}

The brightest parts of the Crab nebula suffer from pile-up in the spectrum. This
is evident from a reduced count rate in observations with 8 CCDs readout as
compared to single CCD readout mode. For a given CCD, the net exposure time per
frame in $k$ CCD readout mode is simply $k$ times higher than the default
integration time of 0.57~s. 

It is difficult to make an exact model for the pile-up. In general, any photon
hitting the detector may give a signal either in a single pixel or in  more than
one pixel, with a range of many possible patterns. Within the same (time) frame,
a second photon may hit the same or a neighbouring pixel, with its own pattern.
The combined number of possible patterns is large, and the electronics, on-board
and ground software treat each combination differently, also depending on the
energy of both photons. 

Therefore we create an empirical pile-up model. We define $\mu$ to be the
expected number of counts per frame per pixel in single CCD readout mode (in the
absence of pile-up). This number is obviously a function of the position on the
detector. Even for sources as bright as Crab, the count rates are moderate. The
maximum value of $\mu$ for the Crab nebula is less than 0.004. Therefore,
already double events are rare and we can fully ignore pixels with three or more
events. We consider only the effect of pile-up on the first-order spectrum. In
general, when a genuine 1st order photon hits a pixel, and within the same frame
a second event (first, second order or even background) hits the same or a
neighbouring pixel, several things may happen but all lead to a rejection of the
event as a first order photon: it is either rejected completely, or seen as a
second order event with twice the energy or alternatively as a background event.
Accounting for pile-up, $\mu_{1,{\mathrm{p}}}$, the expected number of
first-order events in a given pixel can be written as: 
\begin{equation} 
\mu_{1,{\mathrm{p}}} = \mu_1 (1 -ak\mu) 
\label{eqn:pileup} 
\end{equation} 
where $\mu_1$ and $\mu$ are the expected first order and total counts/\ per
pixel per frame without pile-up, $k$ is the number of CCD's read-out and $a$ is
a constant that needs to be determined.

\begin{figure}[!htb]
\begin{center}
\resizebox{\hsize}{!}{\includegraphics[angle=-90]{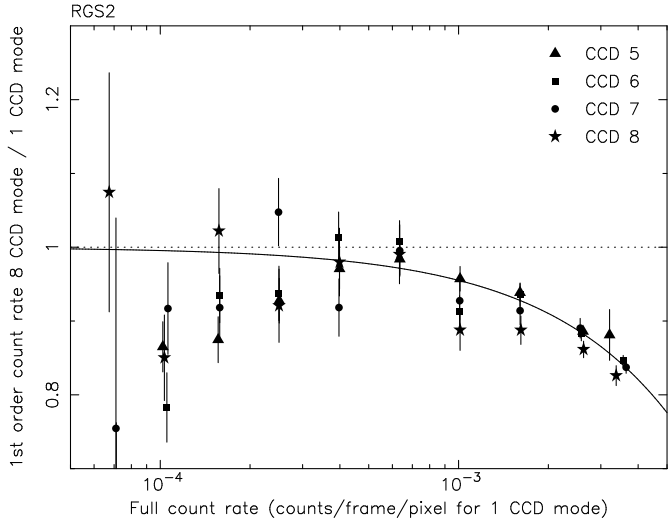}}
\caption{Ratio of first order count rates for the Crab nebula for 8 CCD mode as
compared to single CCD mode, as a function of $\mu$, the total number of counts
per frame per pixel. We only show the results for CCD 5--8 of RGS2. The results
for the other CCDs are similar. The curve indicates our best-fit model to
Eqn.~(\ref{eqn:pileup}).}
\label{fig:pileup}
\end{center}
\end{figure}

For each pixel and all CCDs we have estimated $\mu$ by averaging all counts over
time and $7\times 7$ pixels around the relevant pixel (in order to reduce the
noise). Then we grouped the pixels together in groups with the same range of
$\mu$ values, and determined the total number of first order counts for the
group in each readout mode. The number of first order counts for different $k$,
as a function of $\mu$, were fitted to Eqn.~(\ref{eqn:pileup}). An example is
shown in Fig.~\ref{fig:pileup}. We determined the weighted average of $a$ from
all CCDs. These values are $8.17\pm 1.17$, $6.46\pm 0.30$ and $6.41\pm 0.17$,
from a comparison of $k=1$ with $k=2$, $k=4$ and $k=8$, respectively. Combining
these numbers, our best estimate of $a$ is $6.45\pm 0.15$.  

There is no significant difference between the value of $a$  for RGS1 and RGS2.
As can be seen from Fig.~\ref{fig:pileup}, at low count rates there is a small,
unexplained, deficit of count rate ratio. Fitting the pile-up correction in
Eqn.~(\ref{eqn:pileup}) not to $(1 - ak\mu )$ but to $(b-ak\mu )$, we find that
$b$ is typically a few percent lower than 1. However, as the fit is
statistically driven by the brighter parts with high count rates (see the much
lower error bars for high values of $\mu$), the effect on the derived pile-up
correction at high count rates is small. These brighter parts also dominate the
total count rate of Crab. We estimate that the net effect on the integrated
spectrum is equivalent to an enhanced uncertainty of $\sim 0.45$ in $a$, i.e.
about 3 times the nominal statistical uncertainty in that quantity.

The value for $a$ can also be used for observations of other targets, but it
should be remembered that in our case it is defined for the 98~\% selection in
energy space that we needed to use because of the extended nature of Crab.

The value of $a=6.45$ is used to derive a wavelength-dependent correction
factor for pileup $f_p(\lambda )$, that we formally include in the effective
area correction ("arf"). It is constructed by averaging $1-ak\mu$ over the
cross-dispersion direction, weighted by the observed first order count rate
image corrected for pile-up.

\begin{figure}[!htb]
\begin{center}
\resizebox{\hsize}{!}{\includegraphics[angle=-90]{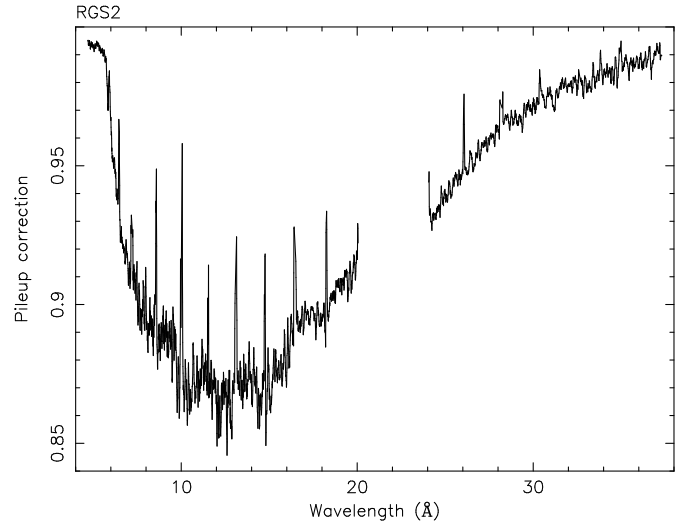}}
\caption{Pile-up correction for the Crab nebula observed with RGS2 in 8 CCD
mode. Most of the spikes of lower pile-up are caused by CCD gaps and the
central columns of CCDs.}
\label{fig:pileuparf}
\end{center}
\end{figure}

In Fig.~\ref{fig:pileuparf} we show the arf correction for the Crab nebula. In
8 CCD mode, the maximum correction occurs at 1~keV and amounts to 14~\%.

\subsection{Vignetting\label{sect:vignetting}}

The effective area of the RGS depends on the position of the source in the field
of view. The standard effective area is calculated for a point source. For
off-axis photons (either in the dispersion or cross-dispersion direction), the
effective area differs due to vignetting of the telescope and obscuration and
vignetting by the gratings. Crab is an extended source. For each energy band we
have calculated the effective vignetting by multiplying a MOS image (see
Sect.~\ref{sect:dispcor}) with the vignetting factor, including both telescope
vignetting and grating obscuration. The average vignetting factor is 0.983, and
to high accuracy it does not depend on energy within the RGS wavelength band.
There is a weak increase of 0.2~\% in telescope vignetting from 7--25~\AA\
caused by the enhanced size of Crab at longer wavelengths, and therefore more
photons are off-axis. But this increase is compensated for by slightly lower
grating obscuration at longer wavelengths. Therefore we multiply our effective
area by this constant factor of 0.983.

\subsection{Cross-dispersion correction\label{sect:crosscor}}

As Crab is an extended source and the RGS detectors have a limited width of
about 5\arcmin\ in the cross-dispersion direction, a fraction of all photons at
each wavelength will not hit the detector. There are three causes for this loss:
the extended emission of the nebula, dust scattering and additional scattering
by the gratings of the RGS. In order to be able to use spectral models for the
full Crab nebula, we need to account for this loss of events. We have
implemented this as follows.

\begin{figure}[!htb]
\begin{center}
\resizebox{\hsize}{!}{\includegraphics[angle=-90]{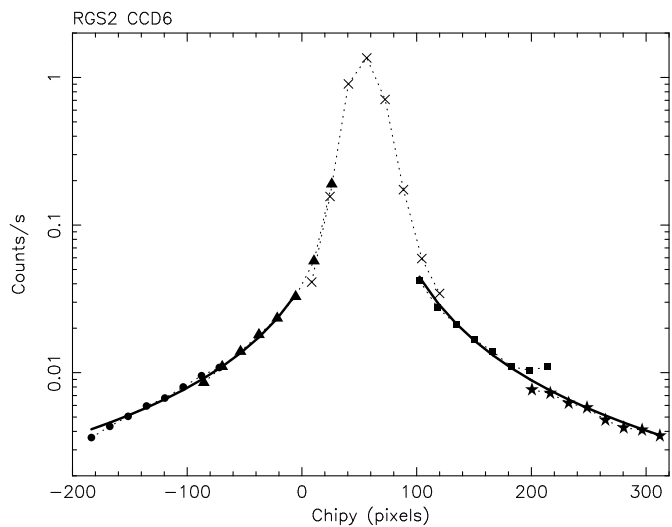}}
\caption{Cross dispersion profile for CCD6 of RGS2 for our 5 observations
(central and two off-axis pointings at each side). The data have been
integrated over the full CCD in the dispersion direction, and over 8 pixels in
the cross-dispersion direction. The best-fit power-law models, fitted
individually to {\it chipy}$<0$ and {\it chipy}$>100$, are shown by the thick
solid lines. Note that we aligned all off-axis pointings to the on-axis ({\it
chipy}) coordinate in the cross-dispersion direction. The different symbols
correspond to the different observations.}
\label{fig:crossprof}
\end{center}
\end{figure}

For our spectral fitting, we only use the on-source pointing, but we use our
off-axis pointings in the cross-dispersion direction to estimate for each
wavelength interval the fraction of the total flux that is lost in the central
pointing. Unfortunately, the extent of the Crab nebula is even larger than our
maximum off-axis angle (9.5\arcmin). Therefore we have to extrapolate the flux
outside this maximum off-axis angle. In general, power-laws are good
approximations for the dust scattering halo's at a given energy
\citep{predehl1995,predehl1996}. Due to the asymmetry of the nebula and the
grating scattering, we determine separate correction factors for both sides of
the nebula. For the longest wavelengths, background becomes important and we
have modelled this by subtracting an observation of the Lockman hole. 

A typical example of our fits is shown in Fig.~\ref{fig:crossprof}. It is seen
that the individual pointings do not fully coincide at their lowest and highest
{\it chipy} positions. This is due to a combination of the outwardly decreasing
flux of the nebula with the slightly asymmetric scattering of the gratings.
However, the total correction factor is hardly affected by this slight lack of
overlap, as we use two full off-axis observations on each side and our fit
averages out these boundary effects.

\begin{figure}[!htb]
\begin{center}
\resizebox{\hsize}{!}{\includegraphics[angle=-90]{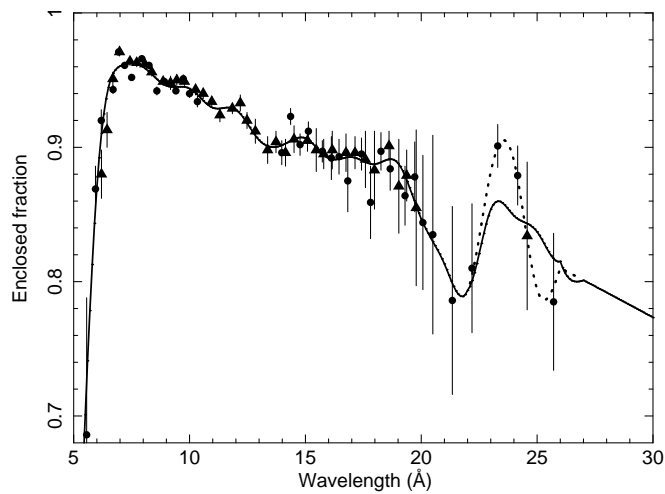}}
\caption{Enclosed flux fraction for the on-axis pointing
on the Crab as a function of wavelength.
Circles are RGS1 data, and triangles RGS2 data, shifted by $+0.02$ in enclosed
fraction. The solid curve shows our finally adopted smoothed approximation to
this correction factor; the dotted line shows the original spline fit to the
data as described in the text.}
\label{fig:enclosed}
\end{center}
\end{figure}

We determined the cross-dispersion loss correction factor $f_{\mathrm d}$
(defined as the fraction of the Crab flux falling on the RGS detector for the
central pointings) with the highest possible resolution. In practice this is
achieved by dividing each CCD into 8 equal wavelength bins. However, in
particular for the CCDs with lower count rates, we had to go to 4 or even 2 bins
per CCD in order to obtain accurate values. We show the measured correction
factors $f_{\mathrm d}$ in Fig.~\ref{fig:enclosed}. The following should be
noted. 

First, for RGS2 $f_{\mathrm d}$ is on average 0.02 lower than for RGS1, caused
by stronger scattering in RGS2. For this reason we have shifted in
Fig.~\ref{fig:enclosed} the RGS2 data points by $+0.02$, in order to eliminate
in first order the differences between both RGS detectors. 

Secondly, the curve shows fine structure near some important edges, such as the
O-K edge (23~\AA), Fe-L (17~\AA) and Mg~K (9~\AA) edges. This structure is real
and is caused by the dust scattering, which is wavelength dependent and shows
structure at these edges (see for example \citealt{draine2003} and
\citealt{costantini2005}). Unfortunately we have insufficient statistics to
increase the wavelength resolution of $f_{\mathrm d}$ near these edges. It is
also hard to model the fine structure theoretically, as it depends on the
chemical and size distribution of the dust grains, the spatial distribution of
the underlying synchrotron emission, and the complex redistribution of photons
in the combined dispersion/cross-dispersion plane of RGS.

Finally, the accuracy of $f_{\mathrm d}$ decreases rapidly towards longer
wavelengths, and beyond 25~\AA\ it is impossible to determine it at all. This is
caused by an accumulation of several effects, namely (1) the decreasing flux of
the Crab nebula towards longer wavelengths due to interstellar absorption, (2)
the stronger dust scattering at long wavelengths, making our power-law fits to
the spatial profile flatter and hence the correction factors higher and more
uncertain, and (3) at the low flux levels of the off-axis pointings at long
wavelengths, systematic background uncertainties emerge.

We approximate the data shown in Fig.~\ref{fig:enclosed} as follows. As a first
approximation, we fit a power-law to the scattered flux fraction, an assumption
that is often made and is based on the Rayleigh-Gans approximation but which
however may not be fully appropriate at the longest wavelengths (see for example
\citealt{smith1998}). The power-law is given by $f_{\mathrm d} =
1-0.00338\lambda^{1.2365}$ with $\lambda$ the wavelength in \AA. A part of this
tail is still visible in our model beyond 27~\AA, and lacking other data, we use
this approximation for all $\lambda >27$~\AA. For the remaining part, we fitted
a spline to the residuals of the power-law fit, and added that spline to our
correction factor. Around 23 and 24~\AA, this spline shows a strong wiggle
caused by the two RGS1 data points at those wavelengths (see the dotted curve in
Fig.~\ref{fig:enclosed}). Such a wiggle is physically not expected and it is
perhaps due to a little overshooting of our fit near the end points. It also
appears that after spectral fitting the predicted flux in the 1--2~\AA\ wide
band just longwards of the O-K edge was too high by up to 4~\%. Therefore we
have fitted a spline to the spectral fit residuals of our preliminary fit in the
23--26~\AA\ range, and included that spline in our final approximation for
$f_{\mathrm d}$, the solid curve of Fig.~\ref{fig:enclosed}. As our spectral
fits show (see Sect.~\ref{sect:specfit}) this gives reasonable results.

\subsection{Broadening in the dispersion direction\label{sect:dispcor}}

The Crab nebula is an extended source. The first order spectrum of any part of
the remnant, at projected distance $\Delta\theta$ (in arcmin) along the RGS
dispersion axis, will be shifted in wavelength by $\Delta\lambda$~(\AA)~$ =
0.138 \Delta\theta$. This holds for all wavelengths. Therefore, the intensity
profile $I(\theta )$ of Crab, integrated over the cross-dispersion direction for
the region covered by the RGS detector, translates directly into a wavelength
broadening profile with the same shape. In our spectral fitting, after deriving
the source spectrum $S(\lambda)$ we convolve this with the broadening kernel
$I(\Delta\lambda)$ before folding with the response matrix. Technically, this is
achieved by using the {\sl lpro} convolution model of the SPEX package, which
has been used successfully for observations with RGS of several extended X-ray
sources like clusters of galaxies, for example \citet{kaastra2001}.

An underlying assumption of this broadening model is that the spatial profile is
independent of wavelength. Our discussion of the Chandra imaging data of M04
(Sect.~\ref{sect:obscurv}) shows that this is strictly speaking not true for the
power-law continuum. However, for the continuum the effect of the broadening is
not important, as it varies on much larger wavelength scales than the broadening
kernel, and M04 also showed that there is little spatial variation in the
interstellar absorption column. Therefore our approximation can be used safely.

\begin{figure}[!htb]
\begin{center}
\resizebox{\hsize}{!}{\includegraphics[angle=-90]{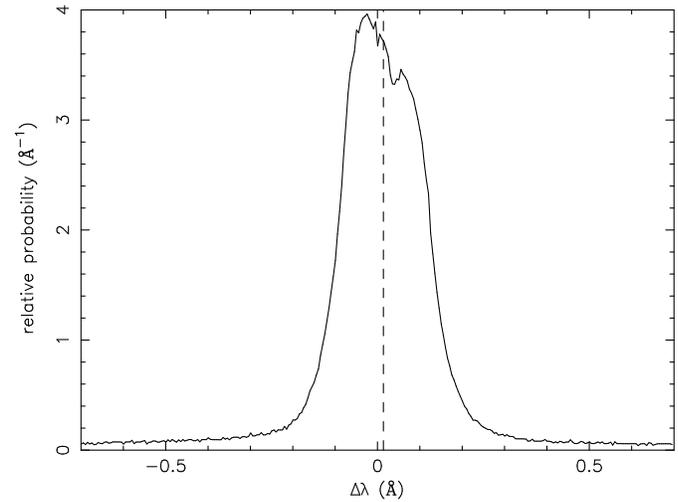}}
\caption{Broadening kernel $I(\Delta\lambda)$ to be used to model the spatial
extent of Crab in the dispersion direction. The kernel is centred around
the nominal position of the pulsar. The average wavelength shift of the kernel
is located at $\Delta\lambda = +0.014$~\AA\ and is indicated with the dashed
line.}
\label{fig:vpro_crab}
\end{center}
\end{figure}

\begin{figure}[!htb]
\begin{center}
\resizebox{\hsize}{!}{\includegraphics[angle=0]{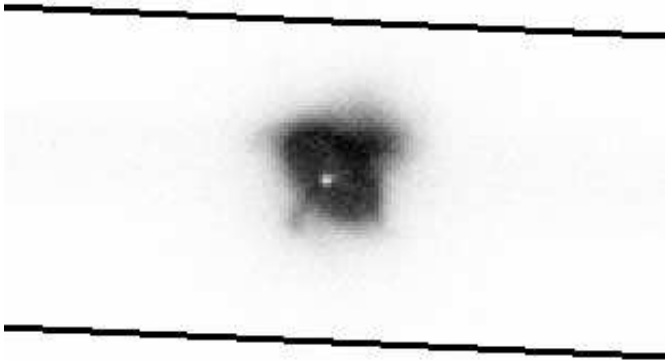}}
\caption{MOS1 image of Crab. Shown is a part of the central CCD image of
observation 0153750501, on a linear intensity scale. North is up. RGS spectra are extracted between the
two parallel lines (separated by 5 arcmin). The pointing is centred on the
pulsar, which shows pile-up in this image.}
\label{fig:mos_crab}
\end{center}
\end{figure}

We have constructed the broadening kernel $I(\Delta\lambda)$ using an XMM-Newton
observation taken in March 2002 (orbit 411, obsid 0153750501). That observation
had the same pointing centre and roll angle as our present observation. MOS1
observed the source in full window mode. The central CCD of MOS1 suffered from
severe loss of events due to telemetry limitations, which are more important
than pile-up for this source. These telemetry losses do not affect the spectral
shape. The remaining fraction of events for the central MOS CCD (3.64~\%) could
be estimated by constructing the projected intensity profiles along each axis
and matching that to the projected profiles of the outer CCDs, that do not
suffer from telemetry saturation or pile-up. Only the pulsar region suffers from
severe pile-up (Fig.~\ref{fig:mos_crab}), but in the RGS band the pulsar contribution is modest (typically
$3-5$~\%). This MOS1 image in the RGS band (0.35--2~keV) was then integrated
over the cross-dispersion of RGS, yielding our kernel $I(\Delta\lambda)$. The
kernel is shown in Fig.~\ref{fig:vpro_crab}. 

The full-width at half maximum of the kernel is 0.22~\AA, which therefore is our
effective spectral resolution. Our spectra were extracted with respect to the
position of the pulsar. The nebula shows a small asymmetry with respect to that
position: the average centre of Crab is shifted by 0.014~\AA.

\subsection{Spectra used for the spectral fitting}

Finally we need to decide which data sets to use. We have spectra in 1, 2, 4 and
8 CCD readout mode. The pile-up corrections increase with the number of CCDs
that are readout simultaneously. As a result, the fits for 8 CCD mode may have
slightly higher systematic uncertainties than the other spectra. Moreover, the
exposure times differ for the readout modes. We have made fits for all
combinations $1$, $1+2$, $1+2+4$ and $1+2+4+8$ CCD readout mode. As a typical
example, the best values for the hydrogen column in units of $10^{25}$~m$^{-2}$
are $3.174\pm 0.036$, $3.175\pm 0.022$, $3.184\pm 0.020$, and $3.160\pm 0.020$
respectively (for model A, without dust). Including the 4 CCD mode data does not
improve the statistics significantly, while including the 8 CCD mode data puts
the best fit parameter near its 1$\sigma$ limit, indicating that the value for
this mode alone deviates more. For these reasons, we here consider only the
combined dataset of single and double CCD readout.

\section{Breakdown of systematic uncertainties}

In this appendix we discuss in more detail the various systematic uncertainties
on our derived effective area solution.

\subsection{Systematic uncertainties on the adopted spectrum of
Crab\label{sect:sys_crab}}

\subsubsection{Systematic uncertainties on the absolute flux level of Crab}

We adopt here the typical uncertainty on the K01 continuum level of 10~\%.
In addition, there is a small uncertainty related to the statistical error
of the slope in the K01 data (Sect.~\ref{sect:intrinsic}). However,
there are also systematic uncertainties in the shape of the K01 continuum.
These are discussed in more detail in the next subsection. We list these
uncertainties as column (a) in Table~\ref{tab:syserr}. 

\subsubsection{Systematic uncertainties on the shape of the Crab spectrum}

The K01 model that we use for the high energy spectrum of Crab is based on
BeppoSAX data. The statistical uncertainties can be determined well
(Sect.~\ref{sect:intrinsic}), but it is harder to determine the systematic
uncertainties.

For the RGS band, the MECS data are most important so we consider those first.
\citet{boella1997} describe in detail the calibration of the MECS, based on
ground measurements and careful modelling of the components of the instrument.
Unfortunately, no estimates of systematic uncertainties are given. However, the
BeppoSAX cookbook \citep{fiore1999} gives some clues. 

First, when the total (nebula plus pulsar) Crab data for MECS, HPGSPC and PDS
are fitted individually, the photon indices derived from the individual
instruments agree to within $\pm 0.016$. 

Secondly, the BeppoSAX cookbook indicates that the average and dispersion of the
PDS to MECS normalisation factor for five different sources excluding Crab, is
0.86 and 0.03, respectively (compare to the value of 0.87 obtained by K01 for
Crab). Adopting this scatter of 0.03 as a typical slope uncertainty over the
effective centres of the MECS and PDS bands at 4.5 and 45~keV, respectively,
we find a corresponding maximum systematic photon index uncertainty of 0.015. 

\begin{figure}
\resizebox{\hsize}{!}{\includegraphics[angle=-90]{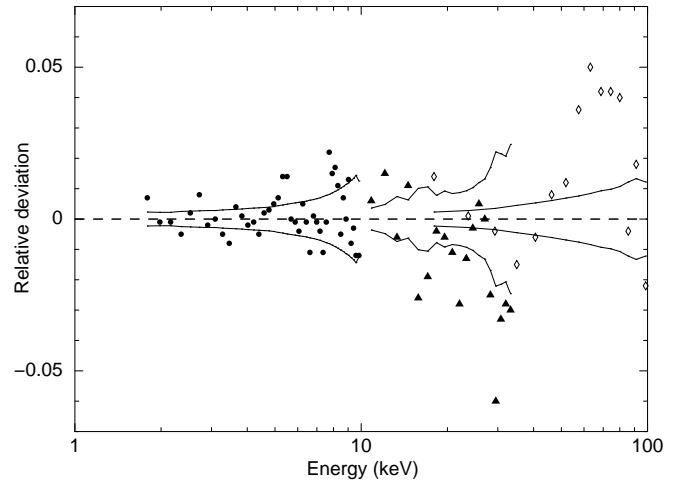}}
\caption{Relative residuals of the BeppoSAX measurements of the Crab nebula used
by \citet{kuiper2001} (Fig.~\protect\ref{fig:crab_k_nebula}) compared to the best
fit model ($\Gamma=2.147\pm 0.002$ and $N=9.31\pm 0.05$). For clarity, error
bars have been omitted, but the solid lines indicate the $\pm1\sigma$
{\sl statistical} error range. Circles: MECS; triangles: HPGSPC; diamonds: PDS. }
\label{fig:kuiper_res}
\end{figure}

Third, we show in Fig.~\ref{fig:kuiper_res} in more detail the residuals of the
fit presented in Fig.~\ref{fig:crab_k_nebula}. These residuals, corresponding to
calibration uncertainties within the instruments, are typically lower than
5~\% and therefore contribute to systematic uncertainties on the photon index 
over the $2-100$~keV band of at most $\pm 0.012$ ($=\log 1.05 / \log 50$).

Finally, the latest calibration of INTEGRAL, covering also a broad energy range
just like the BeppoSAX data used by K01, gives similar results for the hard
X-ray part of the spectrum \citep{jourdain2008}. Down from 3~keV up to the
spectral break at 100~keV, they fit a single power-law with
$\Gamma=2.105\pm0.030$ to the full Crab (nebula plus pulsar) spectrum, using
exactly the same Galactic foreground absorption model as K01. Over the $2-8$~keV
band, the K01 model (curved pulsar plus nebula spectrum) has an effective photon
index of 2.109, only 0.004 higher than the INTEGRAL value. Moreover, the
INTEGRAL flux in this 2--8~keV band is 6~\% higher than the K01 flux in the same
band, well within the quoted 10~\% uncertainty. Thus, if we would have adopted
the INTEGRAL model spectrum instead, the main change would be an overall $\sim
6$~\% higher flux hence a $\sim 6$~\% lower effective area for RGS, but no
significant slope difference. 

The above estimates can be summarised in a maximum systematic slope uncertainty
of about 0.016.

\subsection{Systematic uncertainties on the curvature
correction\label{sect:curcorunc}}

The curvature correction based on Chandra imaging of the nebula as derived in
Sect.~\ref{sect:curcor} has its limitations, leading to systematic uncertainties
in its size. We assess these uncertainties here. 

\subsubsection{Energy range normalisation Chandra continuum}

The choice of the 2--8~keV band to normalise the continuum affects the
correction slightly. Changing the lower limit by 50~\% to 1.5 or 3.0 keV
modifies $f_c(E)$ typically by less than 2.6~\%. We list this as column (b) in
Table~\ref{tab:syserr}.

\subsubsection{Local high-energy cut-offs}

Some of the local power-law fits of M04 may be affected by unnoted high-energy
cut-offs. If this happens, a power-law fit gives a too soft spectrum, resulting
in an over-estimate of $f_c(E)$ at low energies. This effect is hard to
quantify, but the fact that the photon index of $\Gamma=2.10$ derived from a
direct fit to the total ACIS-S spectrum (M04) agrees well with the spatially
integrated spectrum indicates that it is not important. Also, if a significant
fraction of the pixels would have a high-energy cut-off, a clear break in the
BeppoSAX data used by K01 should be present, which is not the case. The maximum
change in photon index for an assumed spectral break near 4~keV in these
BeppoSAX data is less than 0.006. This corresponds to a maximum error in $f_c$
of at most 1.4~\% (column (c)).

\subsubsection{Pile-up corrections Chandra}

M04 needed to make pile-up corrections; they estimate a systematic uncertainty
of 0.05 in their photon indices. As spectral curvature corresponds to
differential photon index, the effect is small (less than 0.03~\%) and we shall
ignore it here. Another potential effect might be a systematic error in the
pile-up correction of M04 that would correlate with intensity and photon index.
The pile-up correction for the photon index $\Gamma$ of M04 is always less than
0.17 for $\Gamma=1.5$ and less than 0.21 for $\Gamma=2.5$. We may assume very
conservatively that the maximum systematic error is half of this difference,
i.e. $0.02$. If we increase and stretch all local photon indices $\Gamma$ with
$\Delta\Gamma (\Gamma) = 0.02 (\Gamma - 2.1)$ and re-evaluate the curvature
correction, we find differences lower than 0.6~\% for $\lambda< 30$~\AA. As
for most wavelength the correction is much lower and the maximum of 0.02 is
likely an over-estimate, we can safely ignore this effect here.

\subsubsection{Different continuum model for different instruments}

In the 1--8~keV band, there is a difference of up to 10~\% between the model
flux of K01 and M04, and a corresponding difference in photon index of
$2.147-2.10=0.047$. Such differences are typical for spectra taken by different
instruments, and generally they correspond to calibration uncertainties. As
above, we do not expect this difference to produce artificial spectral
curvature.

\subsubsection{Interstellar absorption column density}

There is a big difference between the interstellar absorption $N_{\mathrm{H}}$
used in both papers: K01 have $N_{\mathrm{H}}=3.61\times 10^{25}$~m$^{-2}$,
while M04 find $3.20\times 10^{25}$~m$^{-2}$. In the {\sl absorbed} model
continuum near the interstellar oxygen edge at 0.53~keV, this causes a 30/50~\%
higher flux below/above the edge in the M04 model as compared to the K01 model.
Apparently, the difference is caused by a mismatch of the same order of
magnitude between the effective areas of BeppoSAX/LECS and Chandra/ACIS-S near
the oxygen edge. Nevertheless, this does not affect the high-energy continuum,
and it is also not too important for the determination of the spectral
curvature, as that is caused by true spatial variations of the photon index.

\subsubsection{Biased absorption due to region selection}

The value of $N_{\mathrm{H}}$ derived by M04 may be slightly biased, because in
order to avoid pile-up, it was determined in the low surface brightness regions.
These regions are mostly found in the outer parts and have softer intrinsic
spectra. More importantly, the softest X-ray flux in these regions may be
enhanced by up to $\sim 10$~\% due to dust scattering from the brighter central
parts. This may cause a slight under-estimate of the derived interstellar
column, as dust scattering is ignored. On the other hand, the same regions may
loose soft X-ray flux to regions even further out, thereby reducing the effect.
It is therefore hard to quantify this effect, and we will ignore it.

\subsubsection{Spatial variations of $N_{\mathrm{H}}$}

M04 assumed that all parts of the nebula have the same interstellar column
density. This is a reasonable assumption, as they show that less than 0.5~\% of
all pixels has excess absorption up to $7.5\times 10^{24}$~m$^{-2}$ above the
average value. The excess is caused by filaments visible in optical line
emission. The effect of this on the integrated spectrum is less than 0.3~\% near
the oxygen edge. We list this as column (d) in Table~\ref{tab:syserr}.

M04 also estimate that the scatter in $N_{\mathrm{H}}$ due to spatial variations
of the interstellar column density is less than $2\times 10^{24}$~m$^{-2}$. For
the integrated nebular spectrum, most of these variations will average out;
ignoring systematic effects, the uncertainty in the mean absorption for the 2074
pixels considered by M04 would be $\sim\sqrt{2074}$ lower than the scatter,
hence about $4\times 10^{22}$~m$^{-2}$, which is undetectable.

\subsection{Systematic uncertainties in the absorption model}

\subsubsection{The role of dust}

There are small differences between the continuum model derived from model A
(without dust) and model B (with dust). In principle, model B is our preferred
model but it still may have its systematic uncertainties, related to issues as
the exact density, grain chemical composition and size distribution.  As
\citet{seward2006} typically found differences up to 30~\% in the amount of dust
towards Crab depending on their adopted dust model, we will, quite arbitrarily,
take here half of the difference between model A and B as the systematic
uncertainty associated to the dust model. This is listed in column (f) of
Table~\ref{tab:syserr}.

\subsubsection{Hot gas}

The maximum amount of hot gas in the line of sight towards Crab is $N_{\mathrm
H} \la 5\times 10^{23}$~m$^{-2}$ (Sect.~\ref{sect:hotgasdiscussion}). We take
the transmission of gas with such a column density, proto-solar abundances and a
temperature of 0.2~keV as our estimate for the associated systematic uncertainty
introduced by ignoring this component (column (g) of Table~\ref{tab:syserr}).

\subsubsection{The carbon abundance}

We follow here Sect.~\ref{sect:carbon} and adopt a 30~\% systematic uncertainty
in the interstellar carbon abundance. See further column (h) of
Table~\ref{tab:syserr}.

\subsection{systematic uncertainties related to the instrument}

\subsubsection{The cross-dispersion correction}

Our estimate is based on the residuals shown in Fig.~\ref{fig:enclosed} and is
shown in column (i) of Table~\ref{tab:syserr}.

\subsubsection{Pile-up correction}

Our estimate is based on Fig.~\ref{fig:pileuparf}, adopting a 5~\% (2$\sigma$)
uncertainty in the pile-up scale parameter $a$. The uncertainties are shown in
Table~\ref{tab:syserr}, column (j).

\subsubsection{Redistribution errors}

We have calibrated the effective area by multiplying the nominal effective area
by an empirical fudge factor. However, at short wavelengths important
uncertainty is introduced by wide angle scattering of photons by the gratings of
RGS. This cannot be described by a simple arf-correction, as it depends on the
spectral shape of the source. An estimate of the corresponding uncertainty in
the effective area is obtained by comparing the fit residuals for RGS1 and RGS2
separately. We take the difference in residuals, averaged over 1~\AA\ wide bins,
as an estimate for the systematic uncertainty associated to this effect. The
uncertainties are shown in Table~\ref{tab:syserr}, column (k).

\subsubsection{Errors in the Chebychev polynomial corrections}

Our initial effective area was based on observed spectra of Mrk~421; mismatches
between the observed spectra and the adopted power-law continuum of these
sources were corrected using a Chebychev polynomial expansion of a correction
factor for the effective area of RGS (see Sect.~\ref{sect:chebychev}) . We
estimate the uncertainty associated to this by analysing the fit residuals of
another source, \object{PKS~2155-304}, orbit 362. The resulting uncertainties
are shown in column (l) of Table~\ref{tab:syserr}.

\subsubsection{Errors in adopted spatial profile in the dispersion direction}

The Crab spectrum is energy dependent. We have used a MOS1 image in the
$0.35-2.0$~keV range to predict the spatial distribution of the emission in the
dispersion direction of RGS. We have chosen a number of narrow band energy
intervals in the MOS image to see how much that projected profile deviates from
the mean $0.35-2.0$~keV profile. These intervals range from 8 to 22~\AA. We find
shifts lower than 1~m\AA\ in the centroid (to be compared to the FWHM of the
kernel of 220~m\AA), and no deviations higher than 2~\% in the FWHM of the
profile. This justifies our choice of a single broadening kernel. 

\end{appendix}

\end{document}